\documentclass[twocolumn]{aastex61}

\newcommand{\sm}{M_\odot}
\newcommand{\sr}{R_\odot}
\newcommand{\abc}{iPTF\,16abc}
\newcommand{\sneia}{SNe Ia}

\received{2017 August 23}
\revised{2017 November 23}
\accepted{2017 December 5}
\published{2018 January 11}
\submitjournal{ApJ}

\shorttitle{\abc}
\shortauthors{Miller et al.}

\begin{document}

\title{Early Observations of the Type Ia Supernova \abc:\\
A Case of Interaction with Nearby, Unbound Material and/or Strong Ejecta Mixing}

\correspondingauthor{A.~A.~Miller}
\email{amiller@northestern.edu}

\author[0000-0001-9515-478X]{A.~A.~Miller}
\affil{Center for Interdisciplinary Exploration and Research in Astrophysics (CIERA) and Department of Physics and Astronomy, Northwestern University, 2145 Sheridan Road, Evanston, IL 60208, USA}
\affil{The Adler Planetarium, Chicago, IL 60605, USA}

\author[0000-0002-8036-8491]{Y.~Cao}
\affil{eScience Institute and Astronomy Department, University of Washington,
  Seattle, WA 98195}

\author[0000-0001-6806-0673]{A.~L.~Piro}
\affil{The Observatories of the Carnegie Institution for Science, 813 Santa Barbara Street, Pasadena, CA 91101, USA}

\author{N.~Blagorodnova}
\affil{Division of Physics, Mathematics, and Astronomy, California Institute of Technology, Pasadena, CA 91125, USA}

\author{B.~D.~Bue}
\affil{Jet Propulsion Laboratory, California Institute of Technology, Pasadena, CA 91109, USA}

\author[0000-0003-1673-970X]{S.~B.~Cenko}
\affil{NASA Goddard Space Flight Center, Mail Code 661, Greenbelt, MD 20771, USA}
\affil{Joint Space-Science Institute, University of Maryland, College Park, MD 20742, USA}

\author{S.~Dhawan}
\affil{The Oskar Klein Centre, Department of Physics, Stockholm University, AlbaNova, SE-106 91 Stockholm, Sweden}

\author{R.~Ferretti}
\affil{The Oskar Klein Centre, Department of Physics, Stockholm University, AlbaNova, SE-106 91 Stockholm, Sweden}

\author[0000-0003-2238-1572]{O.~D.~Fox}
\affil{Space Telescope Science Institute, 3700 San Martin Drive, Baltimore, MD 21218, USA}

\author{C.~Fremling}
\affil{The Oskar Klein Centre, Department of Astronomy, Stockholm University, AlbaNova, SE-106 91 Stockholm, Sweden}

\author{A.~Goobar}
\affil{The Oskar Klein Centre, Department of Physics, Stockholm University, AlbaNova, SE-106 91 Stockholm, Sweden}

\author[0000-0003-4253-656X]{D.~A.~Howell}
\affil{Las Cumbres Observatory, Goleta, CA 93117, USA}
\affil{Physics Department, University of California, Santa Barbara, CA 93106, USA}

\author[0000-0002-0832-2974]{G.~Hosseinzadeh}
\affil{Las Cumbres Observatory, Goleta, CA 93117, USA}
\affil{Physics Department, University of California, Santa Barbara, CA 93106, USA}

\author[0000-0002-5619-4938]{M.~M.~Kasliwal}
\affil{Division of Physics, Mathematics, and Astronomy, California Institute of Technology, Pasadena, CA 91125, USA}

\author{R.~R.~Laher}
\affil{Infrared Processing and Analysis Center, California Institute of Technology, Pasadena, CA 91125, USA}

\author{R.~Lunnan}
\affil{Division of Physics, Mathematics, and Astronomy, California Institute of Technology, Pasadena, CA 91125, USA}

\author{F.~J.~Masci}
\affil{Infrared Processing and Analysis Center, California Institute of Technology, Pasadena, CA 91125, USA}

\author[0000-0001-5807-7893]{C.~McCully}
\affil{Las Cumbres Observatory, Goleta, CA 93117, USA}
\affil{Physics Department, University of California, Santa Barbara, CA 93106, USA}

\author[0000-0002-3389-0586]{P.~E.~Nugent}
\affil{Lawrence Berkeley National Laboratory, Berkeley, California 94720, USA}
\affil{University of California -- Berkeley, Berkeley, CA 94720, USA}

\author{J.~Sollerman}
\affil{The Oskar Klein Centre, Department of Astronomy, Stockholm University, AlbaNova, SE-106 91 Stockholm, Sweden}

\author{F.~Taddia}
\affil{The Oskar Klein Centre, Department of Astronomy, Stockholm University, AlbaNova, SE-106 91 Stockholm, Sweden}

\author[0000-0001-5390-8563]{S. R. Kulkarni}
\affil{Division of Physics, Mathematics, and Astronomy, California Institute of Technology, Pasadena, CA 91125, USA}



\begin{abstract}

Early observations of Type Ia supernovae (SNe Ia) provide a unique probe of
their progenitor systems and explosion physics. Here we report the
intermediate Palomar Transient Factory (iPTF) discovery of an
extraordinarily young SN Ia, \abc. By fitting a power law to our early light
curve, we infer that first light for the SN, that is when the SN could have
first been detected by our survey, occurred only
$0.15\pm_{0.07}^{0.15}$\,days before our first detection. In the
$\sim$24\,hr after discovery, \abc\ rose by $\sim$2\,mag, featuring a
near-linear rise in flux for $\gtrsim$3\,days. Early spectra show strong
\ion{C}{2} absorption, which disappears after $\sim$7\,days. Unlike the
extensivelyobserved SN Ia SN\,2011fe, the $(B-V)_0$ colors of \abc\ are blue
and nearly constant in the days after explosion. We show that our early
observations of \abc\ cannot be explained by either SN shock breakout and
the associated, subsequent cooling or the SN ejecta colliding with a stellar
companion. Instead, we argue that the early characteristics of \abc,
including (i) the rapid, near-linear rise, (ii) the nonevolving blue colors,
and (iii) the strong \ion{C}{2} absorption, are the result of either ejecta
interaction with nearby, unbound material or vigorous mixing of radioactive
$^{56}$Ni in the SN ejecta, or a combination of the two. In the next few
years, dozens of very young \textit{normal} SNe Ia will be discovered, and
observations similar to those presented here will constrain the white dwarf
explosion mechanism.

\end{abstract}




\section{Introduction}
\label{sec:intro}

Although Type Ia supernovae (SNe Ia) have been extensively used as
standardizable candles, their progenitor systems and explosion physics are
still debated (see a recent review by \citealt{2014ARA&A..52..107M}).
Extremely detailed observations in the hours to days after explosion provide
a promising avenue to further constrain this problem.

While the shock breakout of an SN Ia occurs on a subsecond timescale, the
subsequent quasi-adiabatic expansion and cooling of the unbound ejecta
produces thermal emission that can be used to infer the radius of the
exploding star \citep{2010ApJ...708..598P,2011ApJ...728...63R}. Comparing
models of this cooling emission to the earliest-phase data of SN~2011fe,
\citet{2012ApJ...744L..17B} concluded that the explosion came from a star
with $R_\ast \lesssim 0.02\;\sr$, where $\sr$ is the solar radius. Combining
the radius constraint with the measured ejecta mass,
\citeauthor{2012ApJ...744L..17B} derive the mean density of the progenitor
star, confirming that at least some Type Ia SNe come from compact and
degenerate stars.

Early-phase observations of SNe Ia from a white dwarf (WD)$+$nondegenerate
binary may detect excess emission, relative to most SNe Ia, due to the
collision of the SN ejecta with the nondegenerate companion
\citep{1973ApJ...186.1007W,2010ApJ...708.1025K}. This excess emission was
first detected in iPTF\,14atg \citep{2015Natur.521..328C}, a low-velocity SN
Ia with a significant and declining ultraviolet (UV) pulse detected within a
few days of the SN explosion. This UV pulse is best interpreted as an SN
ejecta--companion collision (but see also
\citealt{2016MNRAS.459.4428K,2017MNRAS.472.2787N}). While such emission
requires a favorable geometric alignment and is only expected in
$\lesssim$10\% of SNe Ia \citep{2010ApJ...708.1025K}, many studies have
searched for signatures of an ejecta--companion interaction, typically
resulting in nondetections (e.g.,
\citealt{2010ApJ...722.1691H,2011ApJ...741...20B,2012ApJ...744...38F,
2012ApJ...744L..17B,2015Natur.521..332O,
2013ApJ...778L..15Z,2015ApJ...799..106G,2016ApJ...826..144S,
2015ApJS..221...22I}). Possible exceptions include SN\,2012cg, which
exhibited excess blue emission in its early-phase light curve
(\citealt{2016ApJ...820...92M}; though for an interpretation that does not
invoke ejecta--companion interaction, see \citealt{2016arXiv161007601S}), and
SN\,2017cbv, which shows a clearly resolved ``bump'' in the early $UBg$
light curves \citep{2017ApJ...845L..11H}.

Interaction is not limited to systems with a nondegenerate companion,
however, as WDs enshrouded in diffuse material following a binary merger
(e.g., \citealt{2015MNRAS.447.2803L}) or expanded owing to a pre-explosion
pulsation can give rise to ejecta-interaction signatures (e.g.,
\citealt{2014MNRAS.441..532D}). Models of this scenario naturally produce
\ion{C}{2} absorption that is comparable in strength to \ion{Si}{2} in the
days after explosion \citep{2014MNRAS.441..532D}, as was observed in
SN\,2013dy \citep{2013ApJ...778L..15Z} and SN\,2017cbv
\citep{2017ApJ...845L..11H}.

The vast majority of SNe Ia are observed to be powered purely by the
radioactive decay of $^{56}$Ni. While the detection of SN shock cooling or
ejecta interaction is rare, the level of $^{56}$Ni mixing in the SN ejecta
can fundamentally alter the appearance of the SN shortly after explosion
(e.g.,
\citealt{2014MNRAS.441..532D,2016ApJ...826...96P,2017MNRAS.472.2787N}).

SNe Ia experience a dark phase after the SN shock breakout but before
radioactive energy diffuses into the photosphere
\citep{2014ApJ...784...85P}. The duration of this dark phase is set by how
the newly synthesized $^{56}$Ni is mixed and deposited into different layers
of the ejecta. Strong mixing leads to a short, or nonexistent, dark phase,
because the radioactive $\gamma$-rays rapidly diffuse to the photosphere.
This also leads to larger luminosities and bluer optical colors at early
times. Even with vigorous mixing it is difficult at very early times, $\ll
1$\,day after explosion, to explain very large luminosities or blue colors,
because the $^{56}$Ni has not had sufficient time to radioactively decay to
$^{56}$Co. If the mixing is weak and the $^{56}$Ni is confined to the
innermost layers of the ejecta, the dark phase can last several days. Weak
mixing results in redder colors and a more moderate rise in luminosity
\citep{2014MNRAS.441..532D,2016ApJ...826...96P}. Thus, the early light
curves of even nonexotic SNe Ia convey information about their progenitor
systems and explosion mechanisms by constraining the distribution of
$^{56}$Ni.

\citet{2017MNRAS.472.2787N} demonstrate that disambiguating between these
different scenarios via optical photometry alone is challenging.
\citeauthor{2017MNRAS.472.2787N} further show that estimates of the time of
explosion, which are critical for comparing models with observations, are
often incorrect by as much as $\sim$2\,days using common methods in the
literature. While analytical models suggest that early spectra can be used to
infer the time of explosion (e.g., \citealt{2014ApJ...784...85P}), more
detailed simulations show that the photospheric evolution is not so simple
\citep{2016ApJ...826...96P}. Reconciling these issues requires both a larger
sample of early SN Ia observations and more detailed models that produce
synthetic light curves and spectra.

In this paper, we report observations of an extraordinarily young SN Ia,
\abc, which was discovered by the intermediate Palomar Transient Factory
(iPTF) on 2016 April $3.36$ UTC at $\textrm{R.A.}=13^h34^m45.49^s$,
$\textrm{Dec.}=+13\degr51\arcmin14\farcs3$ (J2000) with a
$g_\mathrm{PTF}$-band magnitude of $21.44\pm0.25$
\citep{2016ATel.8907....1M}. The transient is spatially coincident with a
tidal tail of the galaxy NGC\,5221, which lies at a distance of
$\sim$100\,Mpc. \abc\ is not detected to a $5\sigma$ limit of
$g_\mathrm{PTF}=21.9$\,mag on April $2.42$, less than 1\,day prior to
discovery, and rose by $\sim$2\,mag in the 24\,hr following its initial
detection. Our spectroscopic follow-up campaign classified \abc\ as a normal
SN Ia \citep{2016ATel.8909....1C}. Our observations and analysis show that
the early evolution of \abc\ exhibited several distinct properties relative
to SN\,2011fe. We interpret those differences as arising from either strong
$^{56}$Ni mixing or ejecta interaction with nearby, unbound material, or a
combination of the two. Alongside this paper, we have released our
open-source analysis and all of the data utilized in this study. These are
available online at \texttt{GitHub}
\url{https://github.com/adamamiller/iPTF16abc}.

\section{Observations}
\label{sec:obs}

During the spring of 2016, the iPTF survey observed the field of \abc\ every
night during dark time in either the $g_\mathrm{PTF}$ or $R_\mathrm{PTF}$
band.\footnote{P48 observations of \abc\ are reported in the
$g_\mathrm{PTF}$ and $R_\mathrm{PTF}$ filters throughout, which are similar
to the SDSS $g'$ and Mould-$R$ filters, respectively (see
\citealt{2012PASP..124..854O} for details on PTF calibration). The
correction from the $g_\mathrm{PTF}$ and $R_\mathrm{PTF}$ filters to SDSS
$g'$ and $r'$ requires knowledge of the intrinsic source color (see
Equations (1) and (2) in \citealt{2012PASP..124..854O}). The spectral
diversity of SNe Ia in the days after explosion is poorly constrained, and
as a result the color terms for \abc\ at these epochs are unknown. We
proceed by assuming that the $g_\mathrm{PTF}$ and $R_\mathrm{PTF}$
calibration is on the AB system, which strictly speaking is incorrect, but
this does not fundamentally alter any of our conclusions.} Survey
observations were conducted with the CFH12K camera
\citep{2008SPIE.7014E..4YR} on the Palomar Observatory 48-inch telescope
(P48; \citealt{2009PASP..121.1395L}). Images were processed by the IPAC
image-subtraction pipeline, which subtracts background galaxy light using
deep pre-SN images and performs forced point-spread function (PSF)
photometry at the location of the SN \citep{2017PASP..129a4002M}. The
photometry is then calibrated to the PTF photometric catalog
\citep{2012PASP..124..854O}.

\begin{figure}[htb] 
    \centering
    \includegraphics[width=3.35in]{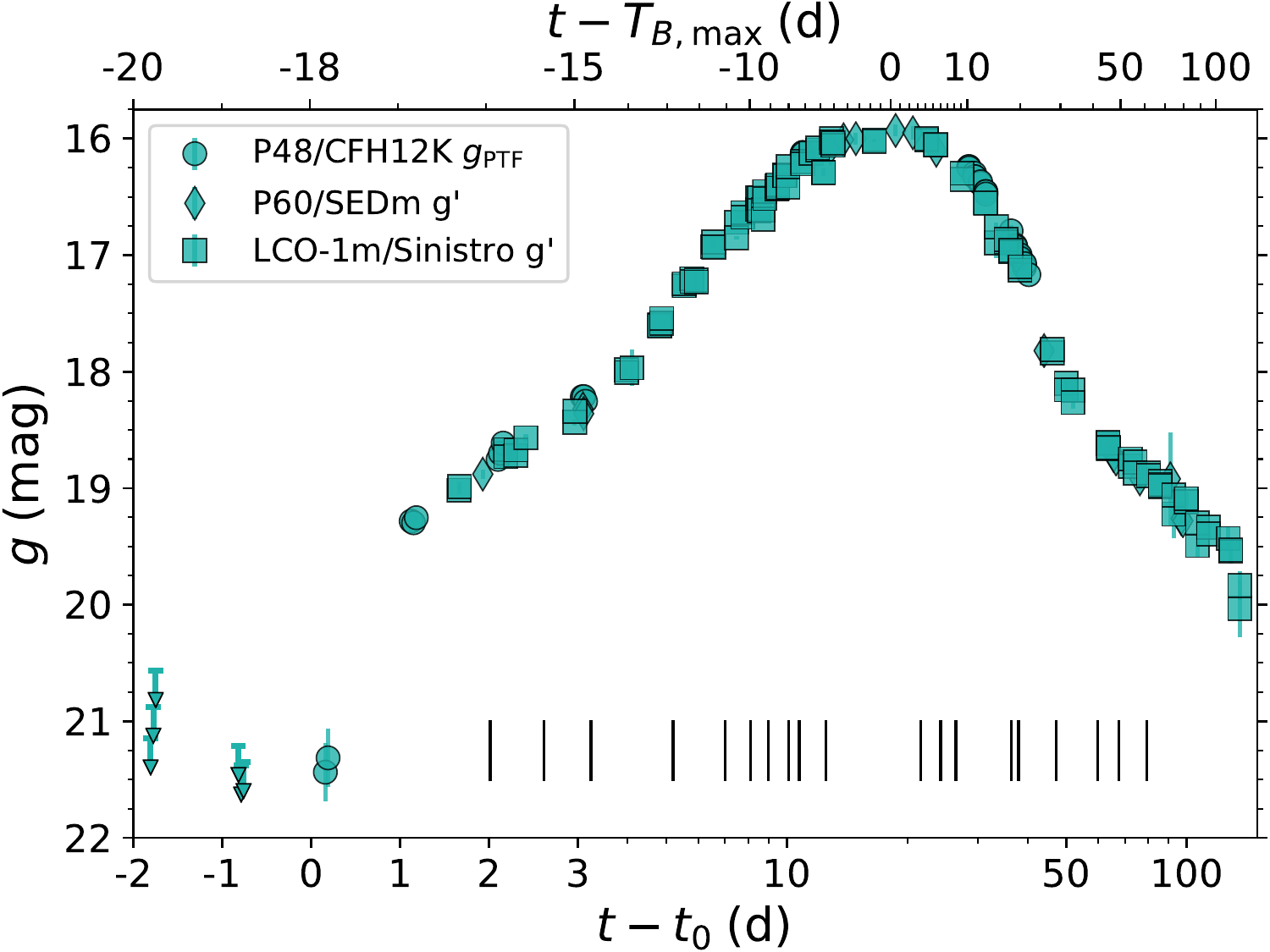} 
    \caption{
    The $g$-band light curve of \abc, with $5\sigma$ upper limits shown as
    downward-pointing arrows. Observations from different telescopes are
    shown with different symbols. The lower axis shows time measured in
    rest-frame days relative to $t_0$ (see \S\ref{sec:lc_fit}), while the
    upper axis shows time relative to $B$-band maximum. Note that the
    horizontal axis is shown with a linear scale for $-2 \; \mathrm{d} \le t
    - t_0 \le 3 \; \mathrm{d}$ and a $\log$ scale for $t - t_0 > 3 \;
    \mathrm{d}$. Vertical black ticks show epochs of spectroscopic
    observations. }
    \label{fig:lightcurve} 
\end{figure}

After discovery, $g'$-, $r'$-, and $i'$-band photometry was obtained with
the SED Machine (SEDm; \citealt{2017arXiv171002917B}) mounted on the Palomar
Observatory 60-inch telescope (P60). We utilized the Fremling Automated
Pipeline (\texttt{FPipe}; \citealt{2016A&A...593A..68F}) to subtract galaxy
light from the SEDm images using archival Sloan Digital Sky Survey (SDSS)
images as a reference. This pipeline then performed forced-PSF photometry at
the location of \abc, which is calibrated to the SDSS catalog
\citep{2014ApJS..211...17A}.

The Las Cumbres Observatory (LCO) 1\,m telescope network obtained $BVg'r'i'$
photometry. PSF photometry was measured on these images using the
\texttt{lcogtsnpipe} pipeline \citep{2016MNRAS.459.3939V}. The \textit{BV}
magnitudes are calibrated to the Fourth USNO CCD Astrograph Catalog
\citep{2013AJ....145...44Z}, and the \textit{g'r'i'} magnitudes are
calibrated to SDSS Data Release 6 \citep{2008ApJS..175..297A}.

The Reionization and Transients InfraRed (RATIR) camera on the autonomous
1.5\,m Harold L. Johnson Telescope
\citep{2012SPIE.8446E..10B,2012SPIE.8444E..5LW} was used to observe \abc\ in
the $r'i'ZYJH$ filters. By design, RATIR lacks a cold shutter, which means
that IR dark frames are not available. Laboratory testing, however, confirms
that the dark current is negligible in both IR detectors
\citep{2012SPIE.8453E..1OF}.

The RATIR data were reduced, co-added, and analyzed using standard CCD and
IR processing techniques in \texttt{IDL}, \texttt{Python},
\texttt{SExtractor} \citep{1996A&AS..117..393B}, and \texttt{SWarp}.
Aperture photometry is obtained following the methods described in
\citet{2014AJ....148....2L}. The $r'i'Z$ filters are calibrated to SDSS
\citep{2014ApJS..211...17A}, while the $JH$ filters are calibrated to the
Two Micron All Sky Survey \citep{2006AJ....131.1163S}. For the $Y$-band
calibration, we used an empirical relation in terms of the $J$ and $H$
magnitudes derived from the United Kingdom Infrared Telescope (UKIRT;
\citealt{2007A&A...467..777C}) Wide Field Camera observations
\citep{2009MNRAS.394..675H}.

The \textit{Swift} satellite observed \abc\ on 14 epochs, beginning
$\sim$15\,days pre-maximum light through $\sim$22\,days post-maximum. The SN
flux is measured via aperture photometry on Ultraviolet-Optical Telescope
(UVOT) images via the usual procedures in \texttt{HEASoft}, including
corrections for coincidence loss and aperture loss. The image counts are
converted to physical fluxes using the latest calibration
\citep{2011AIPC.1358..373B}. There are no pre-SN UVOT images at the SN
location in the \textit{Swift} archive. Visual inspection of the UVOT images
suggests negligible host galaxy contamination in our UVOT flux measurements.
No X-ray emission is detected from \abc\ by the \textit{Swift} X-ray
Telescope (XRT).

The $g$-band discovery and follow-up data of \abc\ are illustrated in
Figure~\ref{fig:lightcurve}. The photometry is shown in the AB system. As
previously noted, the color terms necessary to convert $g_\mathrm{PTF}$ to
the AB system are unknown and assumed to be zero.

\begin{figure*}[!htb]
  \centering
  \includegraphics[width=5.5in]{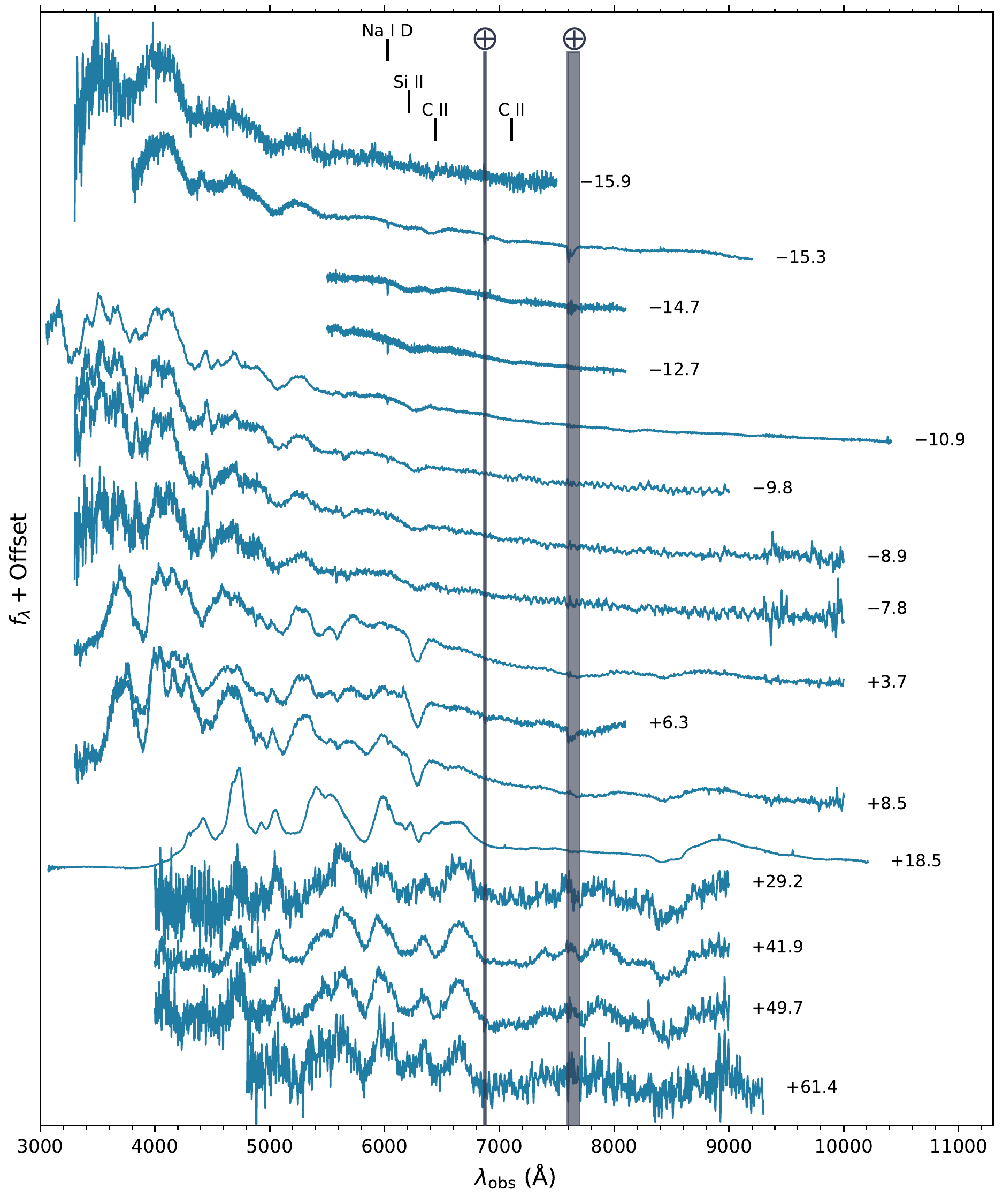}
  \caption{
  Observed spectral sequence of \abc. The spectra are normalized by their
  median flux between 6000 and 7000$\,\textrm{\AA}$. The phase of each
  spectrum relative to the time of $B$-band maximum is shown.
  Telluric absorption bands are grayed out. Line identifications are provided
  for the spectral features discussed in the text. For clarity,
  high-resolution spectra obtained with the VLT have been omitted (see
  \citealt{2017A&A...606A.111F}, for a detailed discussion of these spectra).}
  \label{fig:spec_seq}
\end{figure*}

Spectroscopic observations of \abc\ were taken with a variety of telescopes
and instruments over multiple epochs beginning $\sim$2\,days after discovery
and ending $\sim$2\,months after $B$-band maximum. An observing log is
listed in Table \ref{tab:spec_obs_log}. The spectra were reduced using
standard procedures in \texttt{IDL}/\texttt{Python}/\texttt{Matlab}. The
optical spectral evolution of \abc\ is illustrated in Figure
\ref{fig:spec_seq}, which excludes high-resolution Very Large Telescope
(VLT) spectra for clarity.

\begin{deluxetable}{cccccc}
  \tablecaption{Spectroscopic observations of \abc\ \label{tab:spec_obs_log}}
  \tabletypesize{\footnotesize}
  \tablehead{
    \colhead{Observation} & \colhead{SN} & \colhead{} &
    \colhead{} & \colhead{Range} \\
    \colhead{MJD} & \colhead{Phase} & \colhead{Telescope} &
    \colhead{Instrument} & \colhead{(\AA)}
  }
  \startdata
  $57,483.26$ & $-15.9$ & DCT & DeVeny\tablenotemark{1} & $3301$--$7499$ \\
  $57,483.88$ & $-15.3$ & Gemini-north & GMOS\tablenotemark{2} & $3800$--$9200$ \\
  $57,484.51$ & $-14.7$ & Keck-II & DEIMOS\tablenotemark{3} & $5500$--$8099$ \\
  $57,486.51$ & $-12.7$ & Keck-II & DEIMOS\tablenotemark{3} & $5500$--$8099$ \\
  $57,488.38$ & $-10.9$ & Keck-I & LRIS\tablenotemark{4} & $3055$--$10411$ \\
  $57,489.51$ & $ -9.8$ & LCO-2m & FLOYDS\tablenotemark{5} & $3301$--$8999$ \\
  $57,490.40$ & $ -8.9$ & LCO-2m & FLOYDS\tablenotemark{5} & $3301$--$9999$ \\
  $57,491.55$ & $ -7.8$ & LCO-2m & FLOYDS\tablenotemark{5} & $3300$--$9998$ \\
  $57,492.20$ & $ -7.2$ & VLT & X-shooter\tablenotemark{6} & $3300$--$24550$ \\
  $57,494.00$ & $ -5.4$ & VLT & UVES\tablenotemark{7} & \\
  $57,503.32$ & $ +3.7$ & LCO-2m & FLOYDS\tablenotemark{5} & $3300$--$9999$ \\
  $57,506.00$ & $ +6.3$ & NOT & ALFOSC\tablenotemark{8} & $3602$--$8098$ \\
  $57,508.27$ & $ +8.5$ & LCO-2m & FLOYDS\tablenotemark{5} & $3301$--$9999$ \\
  $57,518.42$ & $+18.5$ & Keck-I & LRIS\tablenotemark{4} & $3071$--$10208$ \\
  $57,520.03$ & $+20.0$ & VLT & X-shooter\tablenotemark{6} & $3300$--$24789$ \\
  $57,529.40$ & $+29.2$ & LCO-2m & FLOYDS\tablenotemark{5} & $4000$--$8998$ \\
  $57,542.41$ & $+41.9$ & LCO-2m & FLOYDS\tablenotemark{5} & $4000$--$8998$ \\
  $57,550.40$ & $+49.7$ & LCO-2m & FLOYDS\tablenotemark{5} & $4001$--$8999$ \\
  $57,562.38$ & $+61.4$ & LCO-2m & FLOYDS\tablenotemark{5} & $4800$--$9300$ \\
  \enddata
  \tablenotetext{1}{The Deveny Spectrograph \citep{2014SPIE.9147E..2NB}}
  \tablenotetext{2}{The Gemini Multi-Object Spectrograph \citep{2004PASP..116..425H}}
  \tablenotetext{3}{DEep Imaging Multi-Object Spectrograph \citep{2003SPIE.4841.1657F}}
  \tablenotetext{4}{Low-Resolution Imaging Spectrometer \citep{1995PASP..107..375O}}
  \tablenotetext{5}{FLOYDS;  \url{https://lco.global/observatory/instruments/floyds}}
  \tablenotetext{6}{X-shooter \citep{XShooter}}
  \tablenotetext{7}{Ultraviolet and Visual Echelle Spectrograph \citep{2000SPIE.4008..534D}}
  \tablenotetext{8}{The Andalucia Faint Object Spectrograph and Camera; \url{http://www.not.iac.es/instruments/alfosc}}
\end{deluxetable}

\section{Host Galaxy, Reddening, and Classification}
\label{sec:usual_staff}

\subsection{Host Galaxy}
\label{sec:host}

\abc\ is spatially coincident with a tidal tail of galaxy NGC\,5221.
\citet{2007A&A...465...71T} derived a distance modulus of
$35.0\pm0.4\,\textrm{mag}$ to NGC\,5221 from the Tully-Fisher relation,
consistent with our derivation from the SN light curve (see
\S\ref{sec:classification}).

Separately, \citet{2015MNRAS.447.1531C} observe the 21\,cm line in NGC\,5221
and measure a redshift of $0.0234$, which we adopt for the remaining
analysis in this paper.

\subsection{Reddening}
\label{sec:reddening}

A detailed study of the reddening toward \abc\ is presented in a companion
paper \citep{2017A&A...606A.111F}. Briefly, the foreground Galactic
extinction toward \abc\ is $E(B-V) = 0.0279$\,mag
\citep{2011ApJ...737..103S}. High-resolution spectra of \abc\ show multiple
absorption components for both the \ion{Ca}{2}\,H$+$K and \ion{Na}{1}\,D
doublets. Despite large equivalent widths (EWs) for these lines, implying
significant extinction (e.g., \citealt{2012MNRAS.426.1465P},
\citeauthor{2017A&A...606A.111F}~find evidence for only a small amount of
extinction. The empirical relation between the EW of \ion{Na}{1}\,D and
extinction has a large scatter, and \citet{2013ApJ...779...38P} have shown
that \ion{Na}{1}\,D absorption is a poor tracer of reddening in SNe Ia.
Thus, we adopt $E(B-V) = 0.05 \, \mathrm{mag}$ as the local extinction for
\abc\ \citep{2017A&A...606A.111F}. For the remainder of our analysis we
assume a total, Galactic$+$host galaxy, line-of-sight extinction of $E(B-V)
= 0.08 \, \mathrm{mag}$.

\subsection{Classification}
\label{sec:classification}

\begin{figure}[htb]
  \centering
  \includegraphics[width=3.35in]{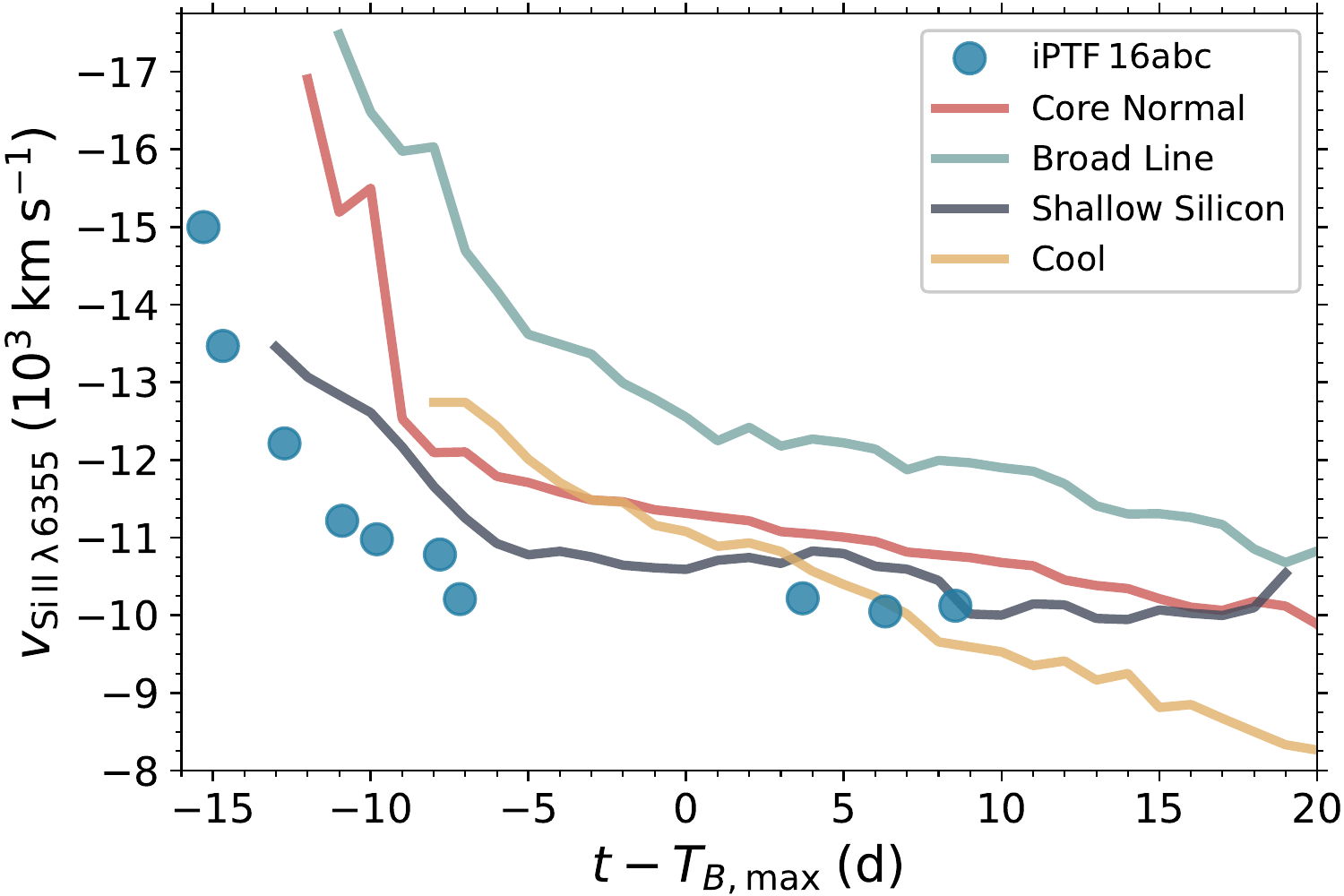}
  \caption{
  Velocity evolution of \ion{Si}{2},
  $v_{\mathrm{Si\,II}\,\lambda6355}$, for \abc\ compared to the median
  evolution of the four spectroscopic subclasses defined in
  \citet{2006PASP..118..560B}, using data from \citet{2012AJ....143..126B}.
  Typical uncertainties for \abc\ are $\sim$1000\,km\,s$^{-1}$ before
  $T_{B,\mathrm{max}}$, when the \ion{Si}{2}$\,\lambda6355$ profile is shallow
  and the minimum of absorption is difficult to determine, and
  $\sim$300\,km\,s$^{-1}$ after $T_{B,\mathrm{max}}$. For the median curves,
  the typical scatter, determined via the interquartile range of the sample,
  is $\sim$700\,km\,s$^{-1}$, around $T_{B,\mathrm{max}}$. At early times,
  core-normal and broad-line SNe have significantly faster \ion{Si}{2} than
  \abc, while the declining trend of cool SNe does not match \abc.}
  \label{fig:branch_vel}
\end{figure}

Using the SuperNova IDentification (\texttt{SNID};
\citealt{2007ApJ...666.1024B}) package, we find that the low-resolution
spectrum of \abc\ at $+18.8$\,days is best matched by normal SNe Ia. Several
characteristic features of an SN Ia, such as \ion{Si}{2} and \ion{S}{2}, can
be easily identified in \abc\ (Figure \ref{fig:spec_seq}). From the $+3.7 \,
\mathrm{day}$ LCO spectrum, we measure the pseudo-equivalent widths (pEWs)
of $-12 \pm 2$\,\AA\ and $-55 \pm 5$\,\AA\ for the absorption features near
5750 and 6100 \AA, attributed to \ion{Si}{2}\,$\lambda\lambda$5972, 6355,
respectively. According to the \citet{2006PASP..118..560B} classification
scheme, \abc\ is a shallow-silicon SN, similar to 1999aa-like SNe
\citep{2009PASP..121..238B}. In Figure~\ref{fig:branch_vel}, the velocity
evolution of the \abc\ \ion{Si}{2}$\,\lambda6355$ absorption minimum is
compared to the median evolution of the four spectroscopic subclasses from
\citet{2006PASP..118..560B}. The median evolution is defined using the
sample of SNe Ia in \citet{2012AJ....143..126B}. For each SN, we interpolate
the \ion{Si}{2} velocity, $v_{\mathrm{Si\,II}\,\lambda6355}$, to a fixed
grid at 1\,day intervals. The curves are defined by the median
$v_{\mathrm{Si\,II}\,\lambda6355}$ at each point on the grid with at least
three SNe (this prevents just one or two SNe from defining the evolution of
an entire subclass). The velocity evolution of \abc\ is most reminiscent of
the shallow-silicon subclass.

To determine the brightness and time of $B$-band maximum for \abc, we fit
the P60 light curves with the \texttt{sncosmo} software
package.\footnote{\texttt{sncosmo} is available at
\url{https://sncosmo.readthedocs.io}.} This fit includes a \texttt{SALT2}
template \citep{2007A&A...466...11G} that has been corrected for extinction
using the \citet{1999PASP..111...63F} reddening law, $R_V=3.1$, and $E(B-V)
= 0.08 \, \mathrm{mag}$.

We determine the time of rest-frame \textit{B}-band maximum to be
$\textrm{MJD}_\mathrm{max}=57499.54\pm0.23$, the coefficient of the zeroth
principle component $x_0 = 0.0086 \pm 0.0003$, the coefficient of the first
principle component $x_1 = 0.96 \pm 0.15$, and the color term $c = 0.033 \pm
0.029$. The best-fit model also gives an unreddened apparent peak magnitude
of $m^*_{B}=15.80 \pm 0.04 \,\textrm{mag}$ in the SN rest frame. In the
following sections, we adopt $\textrm{MJD}_\mathrm{max}=57499.54$ as the
time of $B$-band maximum, $T_{B,_\mathrm{max}}$, and phase $t=0$.

We measure the (pseudo-)bolometric luminosity, $L_\mathrm{UVOIR}$, of \abc\
at peak via trapezoidal integration of the reddening-corrected flux from the
UV, optical, and near-IR (UVOIR) filters. The light curves in the individual
filters are interpolated so that $L_\mathrm{UVOIR}$ is evaluated at common
epochs in each filter. From this integration, we measure a maximum
luminosity $L_\mathrm{max} = 1.2 \pm 0.1 \times 10^{43} \; \mathrm{erg \;
s}^{-1}$ for \abc. This value is consistent with the normal SNe Ia studied
in \citet{2016A&A...588A..84D}. Following Arnett's rule
\citep{1982ApJ...253..785A, 1985Natur.314..337A}, the mass of $^{56}$Ni
synthesized in the explosion can be derived from $L_\mathrm{max}$. Assuming
a rise time of $19 \pm 3$\,days (see \citealt{2006A&A...450..241S}), we find
$M_\mathrm{Ni} = 0.6 \pm 0.1 M_\odot$.\footnote{A 17.9\,day rise time
(\S\ref{sec:lc_fit}), yields a consistent estimate of $M_\mathrm{Ni}$.}

After establishing \abc\ as a normal SN Ia, we use the latest calibration
\citep{2014A&A...568A..22B} of the Phillips relation
\citep{1993ApJ...413L.105P} using $m^*_{B}$, $x_1$ and $c$ to derive a
distance modulus $\mu = 34.89 \pm 0.10 \,\textrm{mag}$ to the SN, provided
that the host galaxy of \abc\ has a stellar mass $< 10^{10}\sm$. A more
massive host galaxy would result in a larger inferred distance modulus that
is nevertheless consistent within the uncertainties. For the following
analysis we adopt a distance modulus $\mu = 34.89 \pm 0.10 \,\textrm{mag}$
for \abc.\footnote{This $\mu$ is consistent with the $z_\mathrm{SN}$, $H_0 =
73 \, \mathrm{km \, s^{-1} \, Mpc}^{-1}$, and Virgo-infall-corrected
distance \citep{2000ApJ...529..786M}.}

\section{Early Observations}
\label{sec:first_light}

Here we consider our suite of early observations of \abc and compare our
findings with SN\,2011fe, a well-studied, nearby SN that was discovered
shortly after explosion
\citep{2011Natur.480..344N,2012ApJ...744L..17B,2014ApJ...784...85P}.

\subsection{Time of First Light from the Early Light Curve}
\label{sec:lc_fit}

The time of first light for SNe is usually estimated by extrapolating
early-phase light curves to determine when the SN flux is equal to 0.
Assuming an ideal, expanding fireball with constant temperature,
\citet{1982ApJ...253..785A} derives that $f \propto t^2$, where $f$ is the
SN flux and $t$ is the time since explosion. Despite these simplified
assumptions, multiple studies have found that the early emission from Type
Ia SNe can be described as a power law in time, with power-law index
consistent with 2, i.e. $f \propto t^2$ (e.g., \citealt{2006AJ....132.1707C,
2010ApJ...712..350H, 2011MNRAS.416.2607G}).\footnote{Many of these studies
sample SN Ia light curves at phases closer to $T_{B,\mathrm{max}}$ than our
initial observations of \abc.}

We model the early flux from \abc\ as a power law:

\begin{equation}
  \label{eq:broken_power_law}
  f(t) \left\{
    \begin{array}{ll}
      = 0,\ \textrm{when}\ t \le t_0 \\
      \propto (t-t_0)^{\alpha},\ \textrm{when}\ t>t_0
    \end{array}
  \right.\ ,
\end{equation}
where $t_0$ is the time of first light, $\alpha$ is the power-law index, and
$t$ is measured in the SN rest frame. We allow $\alpha$ to vary to find the
best match to the data, and we later show that $\alpha = 2$ is not
compatible with the observations. To determine $t_0$ and $\alpha$ we fit the
earliest observations of \abc. Due to slight variations in the passbands,
the model is fit only to the relative $g_\mathrm{PTF}$-band flux.
$g_\mathrm{PTF}$ is the only filter with observations prior to first light,
a necessity for constraining $t_0$.

\begin{figure}[!htb]
  \centering
  \includegraphics[width=3.35in]{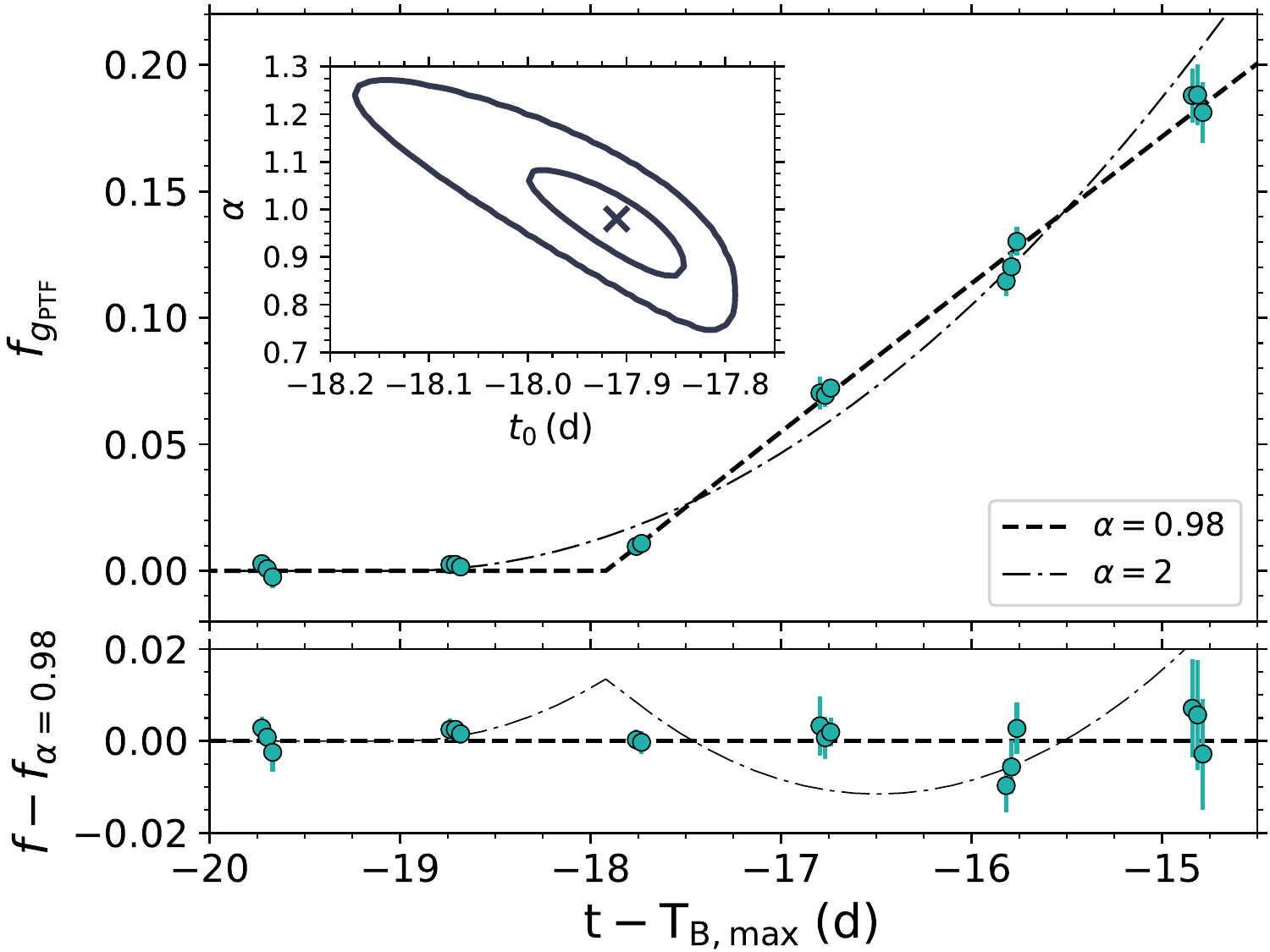}
  \caption{ Best-fit $f \propto t^\alpha$ model to describe the early flux
  from \abc\ in the $g_\mathrm{PTF}$ band. \textit{Top}: The relative flux,
  $f_{g_\mathrm{PTF}}$ (shown as green circles), is measured via forced-PSF
  photometry. The model flux, adopting best-fit parameters $\alpha=0.98$ and
  $t_0=-17.91\,\textrm{d}$, is shown as a thick dashed line. Also shown is
  the best-fit model after fixing $\alpha=2$ (thin dot-dashed line). The
  inset shows the joint distribution of $t_0$ and $\alpha$ for the best-fit
  power-law model. The solid contours represent the $68\%$ and $99.7\%$
  confidence levels. \textit{Bottom}: The observations and models following
  subtraction of the best-fit power-law model, $f_{\alpha = 0.98}$. The
  $t^{0.98}$ model provides a much better fit to the observations than the
  $t^2$ model.}
  \label{fig:early_lc_fit}
\end{figure}

To determine the best-fit parameters, we search a large grid over $t_0$,
$\alpha$, and the proportionality constant and minimize $\chi^2$. The
modeling results show that the SN flux rises approximately linearly between
$t=-18\,\mathrm{d}$ and $t=-15\,\mathrm{d}$. Figure \ref{fig:early_lc_fit}
shows the best-fit result and the joint marginal distribution of $t_0$ and
$\alpha$. From the best-fit model we obtain $\alpha=0.98 \pm ^{0.16}_{0.14}$
and $t_0=-17.91 \pm ^{0.07}_{0.15}$\,days, where the uncertainties represent
the marginalized 95\% confidence intervals. Our first detection of \abc\
occurred $\sim$${0.15}\,\textrm{d}$ after $t_0$. In the analysis that
follows, the precise values of the best-fit parameters are not important.
The critical finding here is that $\alpha \approx 1$ and $t_0 \approx -18 \,
\mathrm{d}$.

Figure \ref{fig:early_lc_fit} also shows the best-fit model while fixing
$\alpha = 2$. The $f \propto t^2$ model does not match the observations.
Formally, for the $\alpha = 2$ model $\chi^2 = 63.7$ with $\nu = 15$ degrees
of freedom (dof), while $\chi^2 = 10.2$ with $\nu = 14$ dof for the $\alpha =
0.98$ model.

As previously noted, a precise determination of the rise time,
$t_\mathrm{rise}$, of SNe Ia is challenging, as there may be a dark phase
following explosion \citep{2014ApJ...784...85P}. Nevertheless, to be
consistent with previous studies (e.g., \citealt{2011MNRAS.416.2607G}), here
we find $t_\mathrm{rise} = 17.91\pm _{0.07}^{0.15}$\,days based on our fit
for $t_0$. We caution, however, that $t_0$ corresponds to the time when
\abc\ was first detectable by P48 and not the time of explosion.

Unlike \abc, the early emission from SN\,2011fe is well fit by an $f \propto
t^2$ model \citep{2011Natur.480..344N}. Thus, the near-linear flux evolution
observed in \abc\ is distinct compared to SN\,2011fe. To our knowledge this
behavior has only been observed in two other SNe (SN\,2013dy and SN\,2014J;
\citealt{2013ApJ...778L..15Z,2014ApJ...783L..24Z,2015ApJ...799..106G}). Any
model to explain the observations of \abc\ must account for this near-linear
rise in the days after first light.

As a brief aside, we note that simulations presented in
\citet{2017MNRAS.472.2787N} show that SN Ia explosion models do not evolve
as a power law in time. \citeauthor{2017MNRAS.472.2787N}\ demonstrate that
$f \propto t^\alpha$ fits to simulated light curves result in glaring errors
to the estimated explosion times. While caution is advised in
\citet{2017MNRAS.472.2787N}, we note that our primary aim with the power-law
fit is to characterize $\alpha$ for \abc\ compared to SN\,2011fe.

\subsection{Time of Explosion from the Photospheric Velocity}
\label{sec:early_vel}

The time of explosion $t_\mathrm{exp}$ is not equal to $t_0$ (see above);
thus, \citet{2014ApJ...784...85P} suggest that measurements of the
photospheric velocity can determine $t_\mathrm{exp}$ given that the ejecta
begin expanding from the moment of explosion. Assuming a constant opacity in
the ejecta, \citeauthor{2014ApJ...784...85P} find that the photospheric
velocity evolves as $v_\mathrm{ph}\propto(t-t_\mathrm{exp})^{-0.22}$.
Numerical experiments by \citet{2016ApJ...826...96P} find that the
constant-opacity assumption strongly depends on the amount of $^{56}$Ni
mixing in the SN ejecta. As a result, the adoption of a $t^{-0.22}$
power-law model may not be valid for all SNe Ia. Nevertheless, we proceed on
the assumption that \abc\ experienced strong $^{56}$Ni mixing (see
\S\ref{sec:Ni_mixing}), corresponding to the models that are best
approximated as a $t^{-0.22}$ power law. We do this in part to compare with
previous studies, though we caution that the inferred value of
$t_\mathrm{exp}$ is subject to uncertainties related to ejecta mixing.

While the photospheric velocity is not easy to measure, line velocities of
\ion{Si}{2} or \ion{Ca}{2} can be used as a proxy
\citep{2014ApJ...784...85P,2016ApJ...826..144S}. In the case of \abc, the
\ion{Ca}{2} IR triplet is very weak, likely due to high temperatures in the
ejecta. Thus, we determine the photospheric velocity from the
\ion{Si}{2}\,$\lambda$6355 line. Visual inspection shows no sign of
multiple-velocity components of \ion{Si}{2}, and that the
\ion{C}{2}\,$\lambda$6580 line overlaps the red wing of the \ion{Si}{2} line
(see Figures~\ref{fig:spec_seq} and \ref{fig:carbon}). Consequently, we model
the observed spectra between 5900 and 6500\,\AA\ (rest frame) as the
combination of two Gaussian kernels plus a linear baseline, which accounts for
\ion{Si}{2}, \ion{C}{2}, and the continuum, respectively. The expansion
velocity of \ion{Si}{2} is measured by the central wavelength of the
\ion{Si}{2} Gaussian kernel.

\begin{figure*}[!thb]
  \centering
  \includegraphics[width=3.35in]{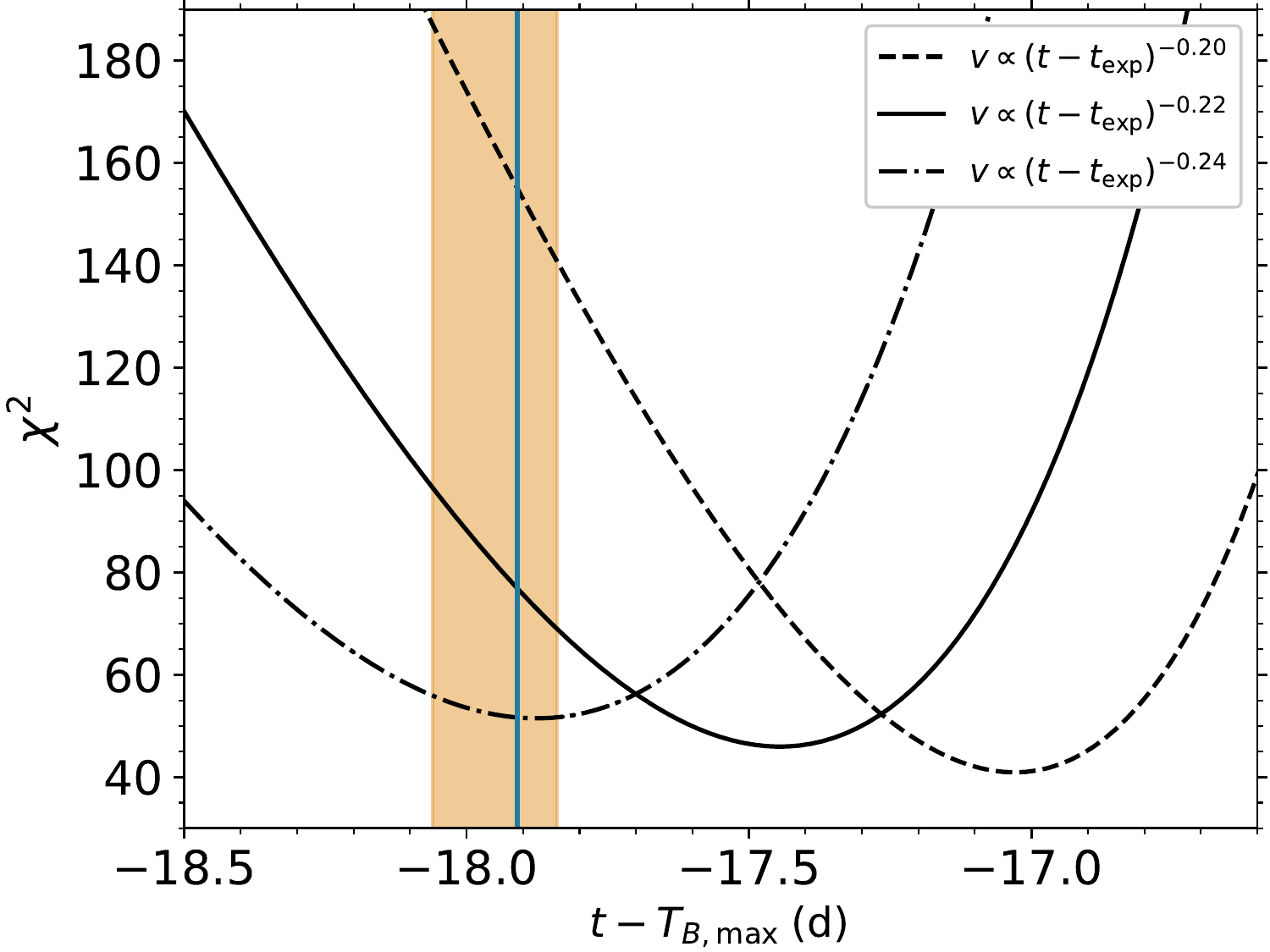}
  \includegraphics[width=3.35in]{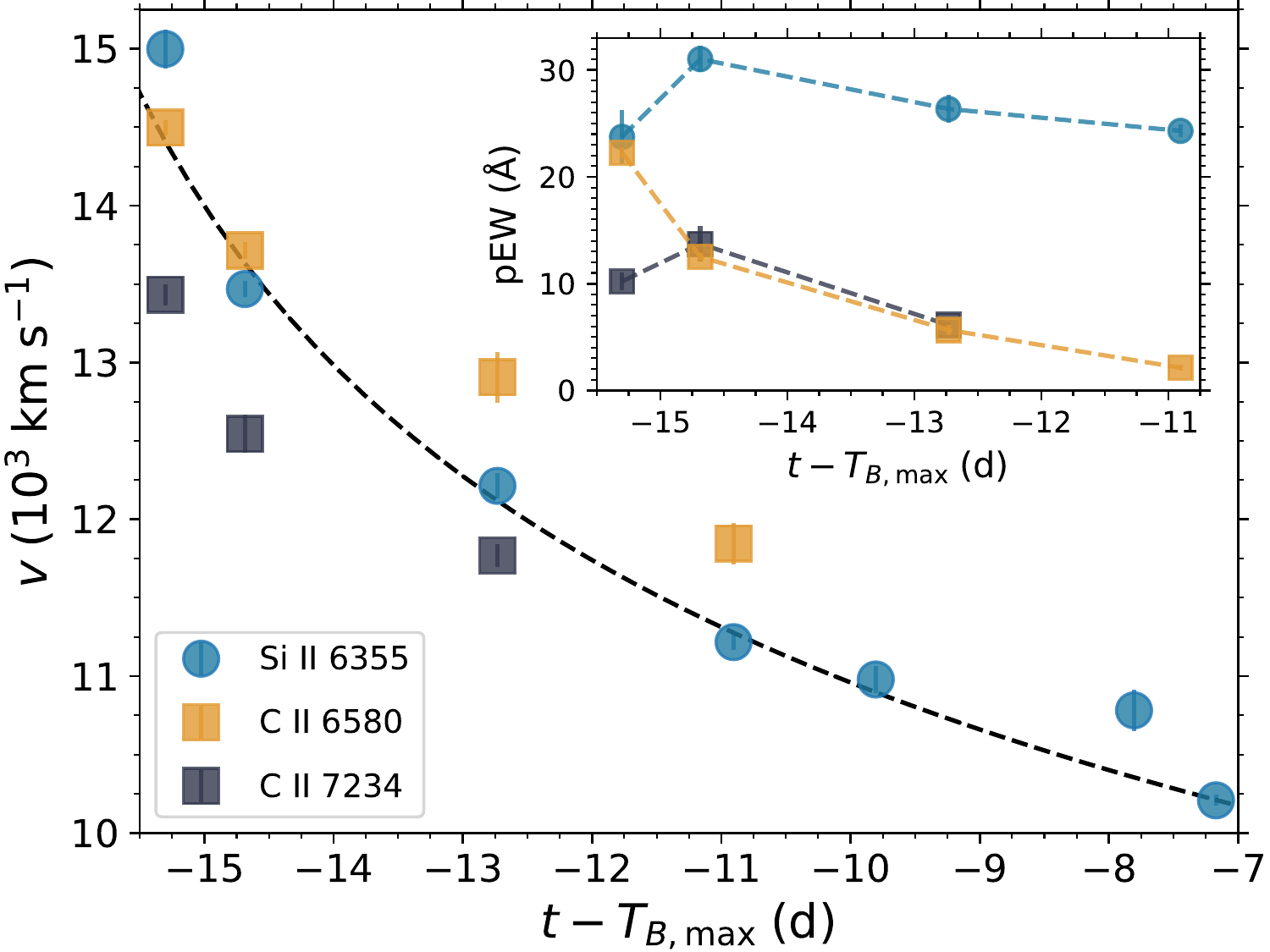}
  \caption{ Constraints on $t_\mathrm{exp}$ from the velocity evolution of
  \ion{Si}{2}. \textit{Left panel:} the dashed, solid and dot-dashed curves
  show $\chi^2$ for fitting power laws with indices $-0.20$, $-0.22$, and
  $-0.24$, respectively. The blue vertical line and the orange shaded region
  indicate $t_0$ and its 95\% confidence interval from Section
  \ref{sec:lc_fit}, respectively. \textit{Right panel:} Observed
  \ion{Si}{2}\,$\lambda$6355 velocities (blue circles) and the best-fit
  power-law model with an index of $-0.22$ (dashed line). For comparison,
  the measured velocities of \ion{C}{2}\,$\lambda\lambda$6580,\,7234 are also
  shown. Typical uncertainties are smaller than the size of the points.
  \textit{Right inset}: vvolution of the pseudo-equivalent width of
  \ion{Si}{2}\,$\lambda$6355 and \ion{C}{2}\,$\lambda\lambda$6580,\,7234 in
  the $\sim$7\,days following explosion.}
  \label{fig:velocity_t_exp}
\end{figure*}

We fit the measured velocities of \ion{Si}{2}\,$\lambda$6355 to the
$v_\mathrm{ph}\propto(t-t_\mathrm{exp})^{-0.22}$ model by minimizing the
$\chi^2$ value and find the best-fit explosion time relative to
$T_{B,\mathrm{max}}$ in the SN rest frame to be $t_\mathrm{exp} = -17.45 \pm
^{0.14}_{0.16}\,\textrm{days}$, where the uncertainties represent the 95\%
confidence interval (Figure \ref{fig:velocity_t_exp}). Following the
analysis in \citet{2014ApJ...784...85P}, we additionally alter the power-law
index to $-0.20$ and $-0.24$ to examine the sensitivity of the result on the
assumed value of $-0.22$. We find that this variation in the power-law index
results in a change of $t_\mathrm{exp}$ of $\approx\pm$0.5\,days (Figure
\ref{fig:velocity_t_exp}). Given the analytical approximation that
$v_\mathrm{ph} \propto t^{-0.22}$, we adopt $t_\mathrm{exp} = -17.5 \pm 0.5
\; \mathrm{d}$, where the uncertainty reflects possible variations in the
power-lax index (see \citealt{2014ApJ...784...85P}).

Comparing our estimates for $t_\mathrm{exp}$ and $t_0$
(Figure~\ref{fig:velocity_t_exp}), we find that $t_0\lesssim
t_\mathrm{exp}$. Since physical causality requires $t_\mathrm{exp} \le t_0$,
we draw the qualitative conclusion that $t_0\simeq t_\mathrm{exp}$, which is
consistent to within the uncertainties. This derivation of $t_\mathrm{exp}$
relies on the assumption $v_\mathrm{ph} \propto t^{-0.22}$, which may not be
valid for all SNe Ia.

\subsection{Strong and Short-lived Carbon Features}
\label{sec:carbon}

\begin{figure}[]
  \centering
  \includegraphics[width=3.35in]{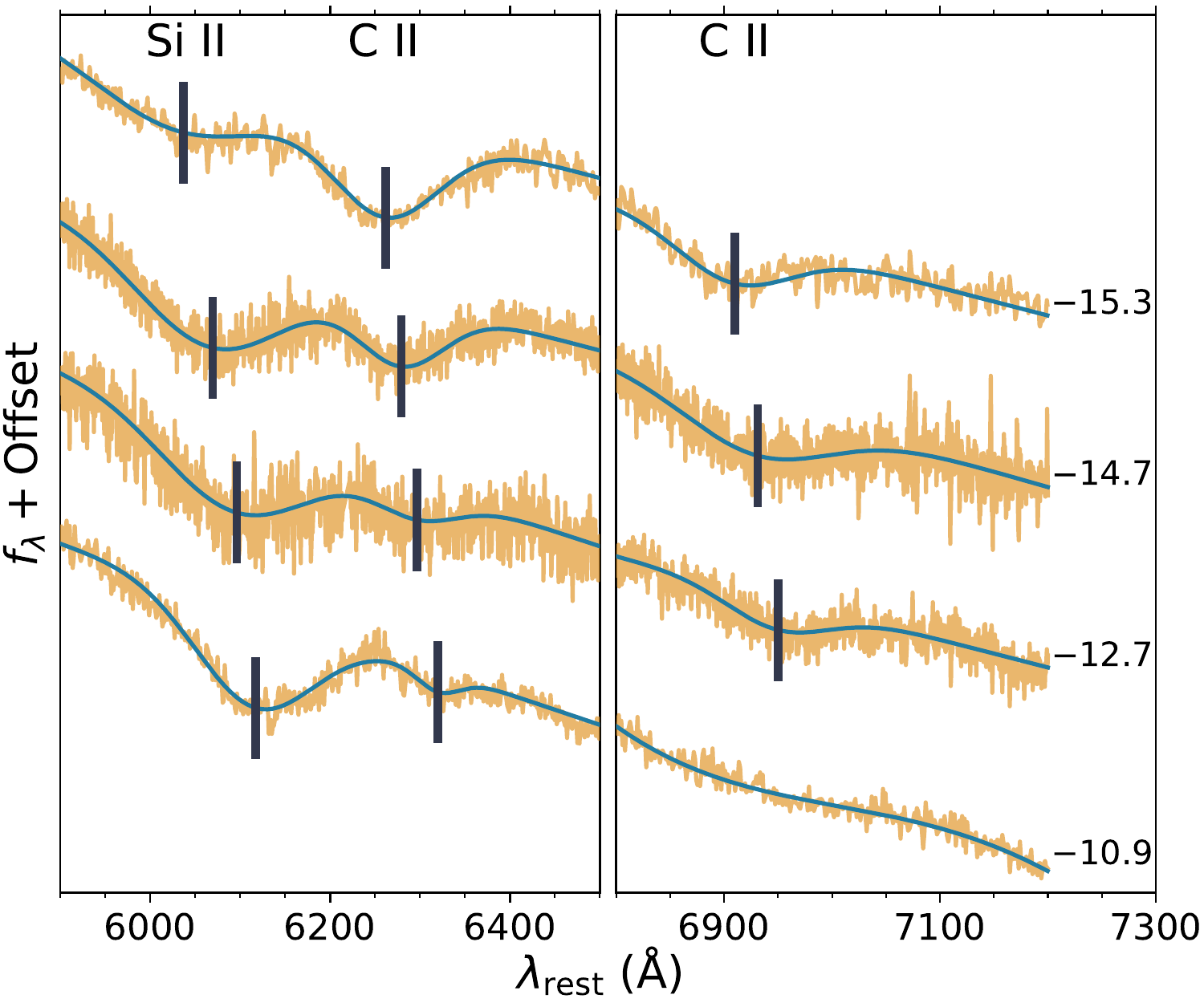}
  \caption{ Evolution of the \ion{C}{2} features observed in the early
  spectra of \abc. The raw spectra are shown in orange, while the solid blue
  lines show the best-fit models (see text for further details). The dark
  gray vertical lines show the measured line centers and clearly show the
  decline in the photosphere velocity in the $\sim$7\,days after explosion (
  \ion{C}{2}\,$\lambda$7234 is not detected in the $-10.9$\,day spectrum).
  The phase of each spectrum relative to $T_{B,\mathrm{max}}$ is labeled. }
  \label{fig:carbon}
\end{figure}

The early spectra of \abc\ exhibit unusually strong \ion{C}{2}
$\lambda\lambda$6580, 7234 absorption. The evolution of these spectral
features is highlighted in Figure~\ref{fig:carbon}, which shows that
\ion{C}{2}\,$\lambda$6580 is as strong as \ion{Si}{2}\,$\lambda$6355 at $t
\approx -15 \, \mathrm{d}$. The strength of the \ion{C}{2} lines declines
with time, and by $t \approx -10 \, \mathrm{days}$ \ion{C}{2} is no longer
detectable.

Similar to our analysis of the \ion{Si}{2}\,$\lambda$6355 line, we can
measure velocities and pEWs of \ion{C}{2}\,$\lambda\lambda$6580, 7234. We
compare the velocity evolution of \ion{C}{2} with that of \ion{Si}{2} in the
right panel of Figure \ref{fig:velocity_t_exp}, which also shows the pEWs of
these lines. These measurements confirm the qualitative analysis from
Figure~\ref{fig:carbon}: namely, the strength of \ion{C}{2}\,$\lambda$6580
is similar to \ion{Si}{2}\,$\lambda$6355 at $t \approx -16 \, \mathrm{d}$
before decreasing and eventually disappearing around $t = -10$\,days.

The detection of \ion{C}{2} in SN Ia spectra is relatively rare, as it
requires both unburned carbon, which is likely only present in the outermost
layers of the ejecta, and nonlocal thermal equilibrium effects in order to
excite the ionized carbon (e.g., \citealt{2007ApJ...654L..53T}). Spectra
obtained around or after $T_{B,_\mathrm{max}}$ rarely show \ion{C}{2} as the
photosphere has receded from the outermost ejecta, while pre-maximum spectra
show evidence for weak \ion{C}{2} absorption in $\sim$1/4 of all normal SNe
Ia (e.g.,
\citealt{2011ApJ...732...30P,2011ApJ...743...27T,2012MNRAS.425.1917S}).
While the sample of SNe Ia with spectra taken within a few days of explosion
is small, SN\,2013dy and SN\,2017cbv are the only other objects known to
have strong \ion{C}{2} features like \abc\
\citep{2013ApJ...778L..15Z,2017ApJ...845L..11H}. As a counterexample,
SN\,2011fe only exhibited weak \ion{C}{2} features in its first spectra
\citep{2012ApJ...752L..26P}. Thus, models of \abc\ must explain the strong
\ion{C}{2} absorption observed shortly after explosion.

\subsection{Blue Optical Colors Shortly after Explosion}

\begin{figure}[]
  \centering
  \includegraphics[width=3.35in]{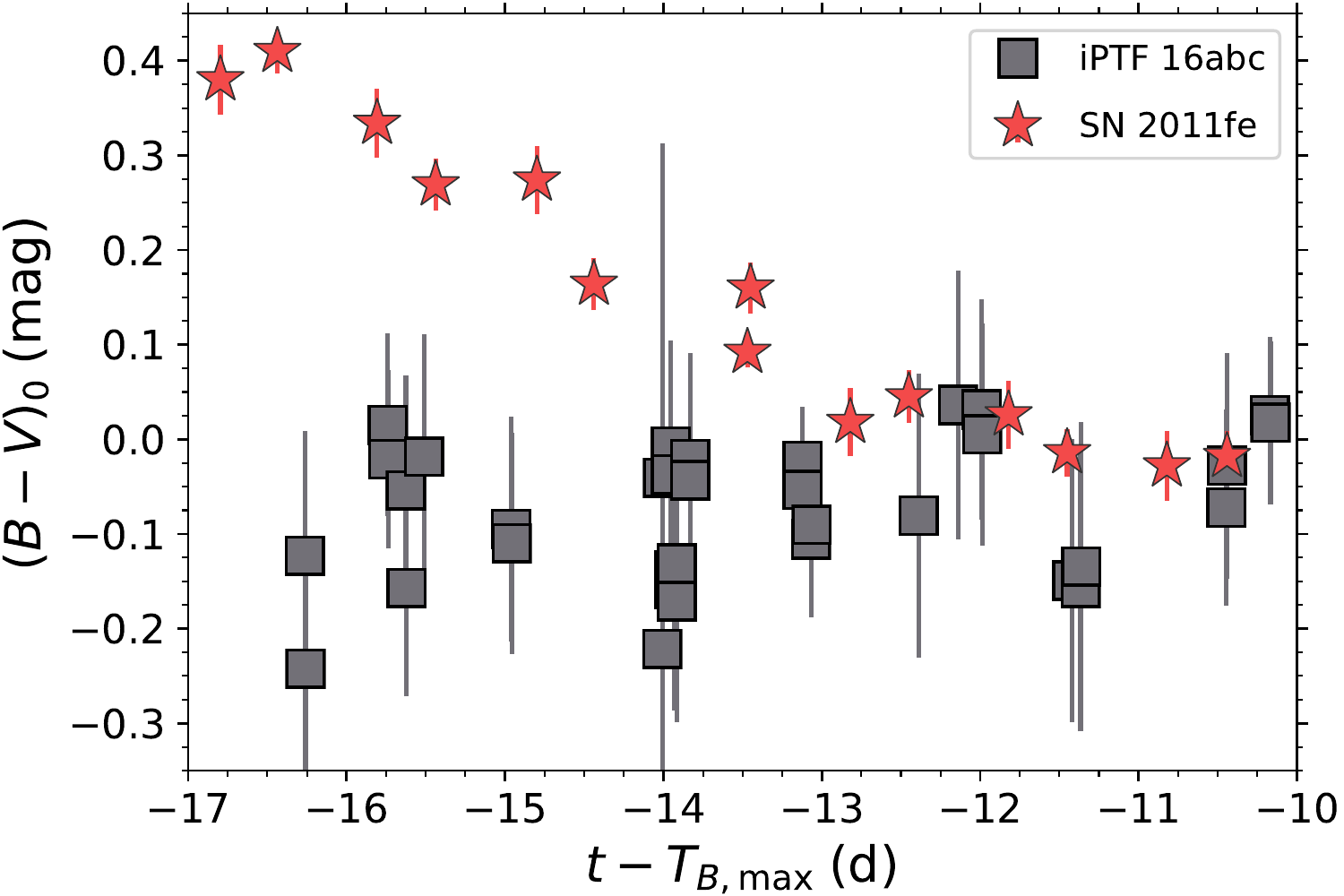}
  \caption{$(B - V)_0$ color evolution of \abc\ (squares) 
    compared to SN\,2011fe (stars). The $B$ and $V$ photometry are calibrated
    on the Vega system, and have been corrected for extinction. The data for
    SN\,2011fe are from \citet{2016ApJ...820...67Z}.}
  \label{fig:B-Vcolors}
\end{figure}

Multiband observations of \abc\ began $\sim$1.5\,days after discovery, which
allows us to trace its color evolution starting $\sim$1.7\,days after $t_0$.
In Figure~\ref{fig:B-Vcolors} we compare the $(B - V)_0$ color evolution of
\abc\ to observations of SN\,2011fe \citep{2016ApJ...820...67Z}. For both
SNe the colors have been corrected for the total inferred reddening along
the line of sight. Interestingly, \abc\ has a nearly flat color evolution up
to $t \approx -10$\,days, while SN\,2011fe initially exhibits red colors
before evolving to the blue.

Roughly 16\,days prior to $T_{B,_\mathrm{max}}$, the $(B - V)_0$ color of
\abc\ is $\sim$0.5\,mag bluer than SN\,2011fe. Like \abc, SN\,2012cg
\citep{2016ApJ...820...92M} and SN\,2017cbv \citep{2017ApJ...845L..11H}
exhibit $(B - V)_0$ colors that are significantly bluer than SN\,2011fe at
very early epochs. While there are many factors that contribute to the early
optical colors of SNe Ia (see \S\ref{sec:lc_energy} below), early blue
colors are often interpreted as a hallmark of interaction between the SN
ejecta and a binary companion. Nevertheless, it is interesting to note that
despite the blue optical colors, the $\mathrm{UV} - \mathrm{optical}$ colors
of \abc, SN\,2012cg, and SN\,2017cbv are significantly redder at these early
epochs than the $\mathrm{UV} - \mathrm{optical}$ colors of iPTF\,14atg
\citep{2015Natur.521..328C}, the most likely candidate for SN
ejecta--companion interaction.

\section{Modeling the Early Evolution of \abc} \label{sec:lc_energy}

Relative to the nearby, normal SN\,2011fe, we have identified several
distinct characteristics of the early evolution of \abc, including (i) a
near-linear photometric rise; (ii) a qualitatively short, or possibly
absent, dark phase, assuming $v_\mathrm{ph} \propto t^{-0.22}$; (iii) the
presence of strong \ion{C}{2} absorption; and (iv) blue and nearly constant
$(B - V)_0$ color in the week after explosion. While most SNe Ia are powered
purely by radioactive decay, the observed radiation shortly after explosion
can also include contributions from SN shock cooling or the collision of the
SN ejecta with a nondegenerate companion or nearby, unbound material. Here
we consider these scenarios as possible explanations for the early behavior
of \abc.

\subsection{SN Shock Cooling}

The shock breakout of an SN Ia lasts for a fraction of a second due to compact
size of the exploding star. Emission from the subsequent cooling phase may
last for several days, however (e.g., \citealt{2010ApJ...708..598P}).
Following the analysis of \citet{2012ApJ...744L..17B} for SN\,2011fe, we
compare the early-phase $g_\mathrm{PTF}$ light curve of \abc\ with two shock
cooling models \citep{2010ApJ...708..598P, 2011ApJ...728...63R}. From this
analysis, we constrain the \abc\ progenitor radius to be $<1\,\sr$. Our
observations of \abc\ cannot place tight constraints on the size of its
progenitor. Indeed, for a typical WD radius, such as that inferred for
SN\,2011fe ($\lesssim 0.02$--$0.04\,\sr$; \citealt{2012ApJ...744L..17B,
2014ApJ...784...85P}), the expected emission from shock cooling is
$\sim$2\,mag fainter than the P48 $g_\mathrm{PTF}$ detection limit at this
distance. Thus, we conclude that shock cooling does not contribute to the
early emission detected from \abc.

\subsection{SN--Companion Collision}
\label{sec:companion}

The detection of emission from the collision of the SN ejecta with a
nondegenerate companion requires a favorable orbital alignment relative to
the line of sight. Thus, from geometric considerations alone the probability
of detecting ejecta--companion interaction is low, $\sim$10\%.
\citet{2010ApJ...708.1025K} calculates that the collision of SN ejecta with
a companion generates thermal emission with a spectrum that peaks in the UV.
The resulting \textit{g}-band emission is expected to be weak.

\begin{figure}[!thb]
  \centering
  \includegraphics[width=3.35in]{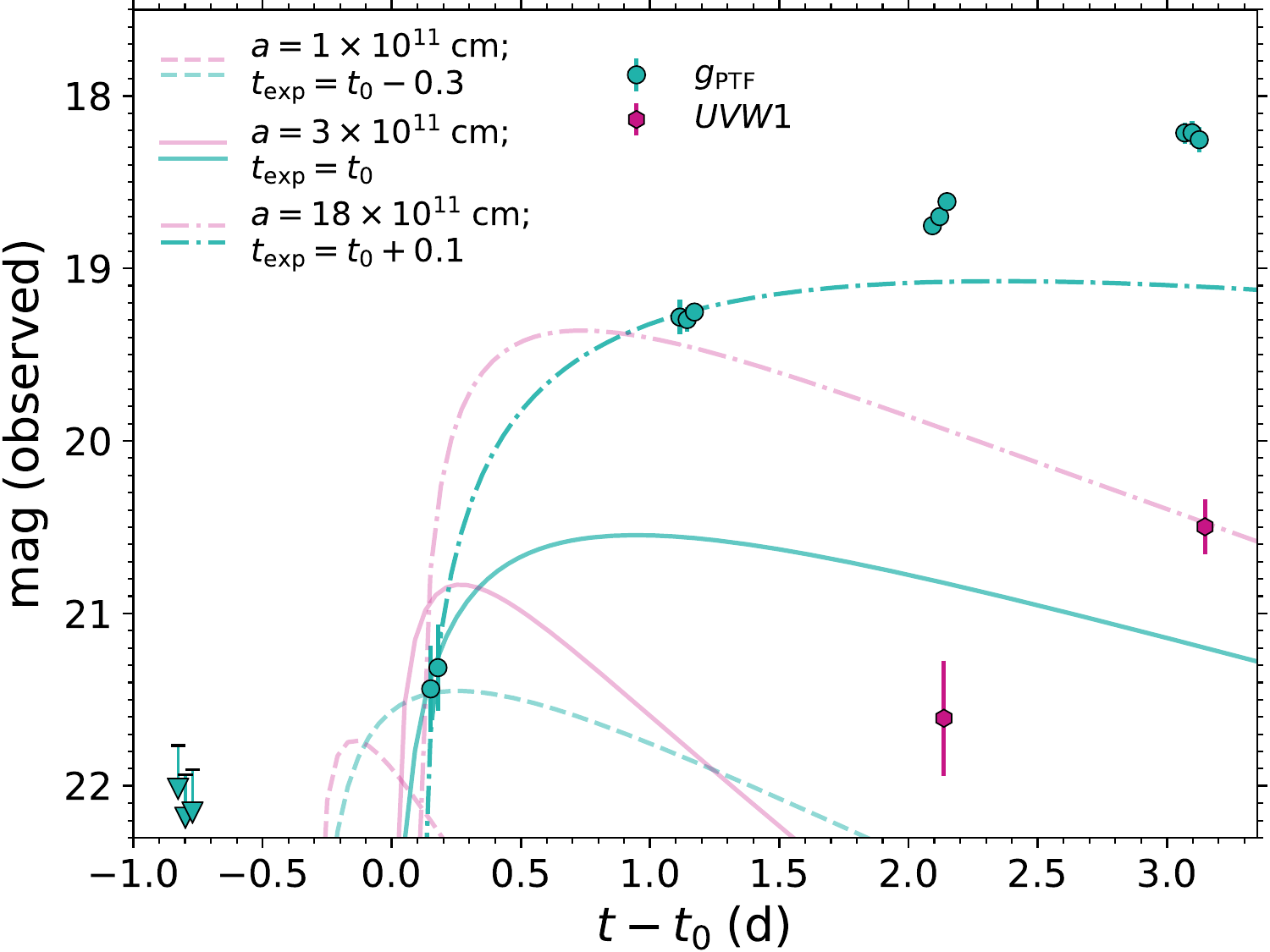}
  \caption{Comparison of SN ejecta--companion interaction models with early
  observations of \abc. $g_\mathrm{PTF}$ detections and 3$\sigma$ upper
  limits are shown as green circles and downward-pointing arrows,
  respectively. \textit{Swift}/$UVW1$ observations are shown as magenta
  hexagons. The dashed, solid, and dot-dashed lines show the expected flux
  for companion-interaction models in the $g_\mathrm{PTF}$ (green) and
  $UVW1$ (magenta) filters. The models have been adjusted to account for the
  distance and reddening toward \abc. Each model features a different
  companion semi-major axis, $a$, and time of explosion, $t_\mathrm{exp}$,
  as labeled in the legend. While models with $a \gtrsim 10^{12} \,
  \mathrm{cm}$ can explain the early optical rise, they greatly overpredict
  the UV flux. Models with $a \approx 10^{11} \, \mathrm{cm}$ can explain
  the initial detection of \abc, but they fail to replicate the $\sim$2\,mag
  rise in the $\sim$24\,hr after explosion. }
  \label{fig:SN--companion}
\end{figure}

To examine the possibility of a SN--companion signature in the early light
curve of \abc, we employ the \citet{2010ApJ...708.1025K} model and assume
canonical values for the ejecta mass, $1.4\,\sm$, expansion velocity,
$10^{4}\,\textrm{km}\,\textrm{s}^{-1}$, and a constant opacity,
$0.2\,\textrm{cm}^2\,\textrm{g}^{-1}$. We calculate the expected
$g_\mathrm{PTF}$ brightness of an ejecta--companion collision at the
distance of \abc\ behind a total reddening of $E(B-V) = 0.08 \,
\mathrm{mag}$ using the parameterized equations in
\citet{2012ApJ...749...18B}. If we assume that the binary is aligned with
the optimal orientation relative to the line of sight, a binary separation
of $a \approx 3 \times 10^{11}\, \mathrm{cm}$ is needed to explain the
initial detection of \abc, as shown in Figure~\ref{fig:SN--companion}. The
minimum binary separation capable of explaining the observed brightness at
the epoch of discovery is $a \approx 10^{11} \, \mathrm{cm}$.
Figure~\ref{fig:SN--companion} shows that such models peak at
$g_\mathrm{PTF} \approx 21.5 \, \mathrm{mag}$, provided that $t_\mathrm{exp}
\approx t_0 - 0.3 \, \mathrm{days}$. These models do not, however, match the
$g_\mathrm{PTF}$ evolution for $t > t_0 + 0.5 \, \mathrm{days}$ (though it
is possible that the SN photosphere dominates the companion-interaction
signature at this phase). While they are otherwise compatible with the
observations, we do not favor the above models as the explanation for the
early flux from \abc\ because they do not explain the 2\,mag rise in the
$\sim$24 hr after discovery.

Figure~\ref{fig:SN--companion} additionally shows that a companion at $a
\approx 18 \times 10^{11} \, \mathrm{cm}$ provides a good match to the
initial optical rise, if $t_\mathrm{exp} \approx t_0 + 0.1 \;
\mathrm{days}$. Models with $a \gtrsim 10^{12} \, \mathrm{cm}$, which can
explain the initial $g_\mathrm{PTF}$ rise, significantly overpredict the
observed UV flux, however. There is no choice of $a$ capable of replicating
the early rise of \abc\ without also overpredicting the observed UV flux.

The challenges associated with each of the previously considered models lead
us to conclude that the early evolution of \abc\ cannot be explained via
ejecta--companion interaction. We cannot, however, exclude the presence of a
red giant, or other nondegenerate, companion as our calculations have
assumed that the binary is aligned with the optimal geometry relative to the
line of sight. If the geometry is not favorable, then it is possible that
signatures from interaction with a companion are not visible.

\subsection{Sub-Chandrasekhar Detonations and Pure Deflagrations}

In \citet{2017MNRAS.472.2787N} the early photometric evolution of SNe Ia is
explored via a variety of explosion models and detailed radiative transfer
calculations. Specifically, \citet{2017MNRAS.472.2787N} examine two
Chandrasekhar-mass ($M_\mathrm{Ch}$) explosions and compare their evolution
to sub-Chandrasekhar detonations and pure deflagrations. The $M_\mathrm{Ch}$
explosions include the ``W7'' carbon-deflagration model of
\citet{1984ApJ...286..644N} and the ``N100'' delayed-detonation model from
\citet{2013MNRAS.429.1156S}. The sub-Chandrasekhar models include a violent
WD-WD merger, which triggers a carbon detonation in the more massive WD, a
centrally ignited sub-Chandrasehkar detonation, and a sub-Chandrasehkar
double-detonation explosion, in which an He-surface-layer detonation
triggers a carbon detonation in the core. \citet{2017MNRAS.472.2787N} note
that, of these last two sub-Chandrasehkar models, the latter provides the
more realistic scenario. Finally, \citeauthor{2017MNRAS.472.2787N} also
examine the ``N5def'' and ``N1600Cdef'' pure deflagration explosions from
\citet{2014MNRAS.438.1762F}. The ``N5def'' model is particularly unique in
that the explosion does not fully unbind the WD, meaning that, unlike with
typical SNe Ia, a remnant remains.

\begin{figure}[]
  \centering
  \includegraphics[width=3.35in]{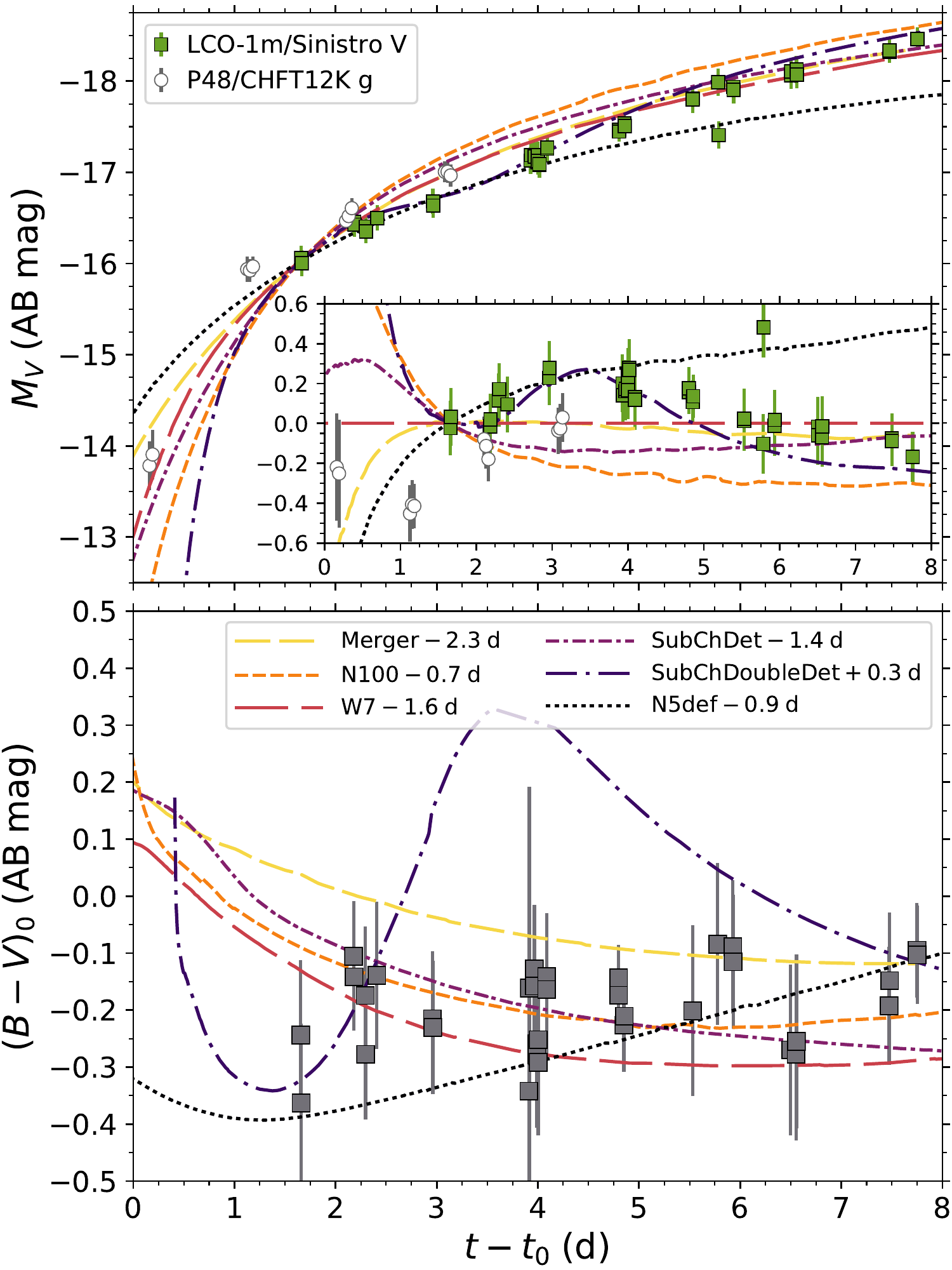}
  \caption{
  Comparison of the models from \citet{2017MNRAS.472.2787N} to the \abc\
  $V$-band light curve (\textit{top}) and the $(B-V)_0$ color curve
  (\textit{bottom}). The light curve has been corrected for the distance
  modulus to \abc\ (\S\ref{sec:host}), while both the light curve and color
  curve have been corrected for reddening. To guide the eye,
  $g_\mathrm{PTF}$ observations are shown, though we caution that \abc\ may
  exhibit significant color evolution at this phase, in which case
  $g_\mathrm{PTF}$ would be a poor proxy for $V$. Each model light curve is
  translated to match the LCO observations $\sim$1.7\,days after $t_0$ given
  the uncertain time of explosion. Translational offsets are listed in the
  bottom panel legend, which shows the models in order of decreasing
  $^{56}$Ni from top to bottom, then left to right. The inset in the top
  panel shows the residuals relative to the W7 model.}
  \label{fig:noebauer} 
\end{figure}

In Figure~\ref{fig:noebauer} we compare photometric observations of \abc\ to
the models presented in \citet{2017MNRAS.472.2787N}.\footnote{Available at
\url{https://hesma.h-its.org/}.} Interestingly, the sub-Chandrasekhar
double-detonation model (subChDoubleDet) replicates the early wiggle in the
$V$-band light curve. However, this match requires an explosion
\textit{after} our initial detection of \abc and predicts extreme color
evolution that is not observed. Thus, the subChDoubleDet model is
incompatible with the observations. The sub-Chandrasekhar detonation
(subChDet) provides a better match to the observations, though this model is
not favored as a particularly realistic scenario (see above). Of the
sub-Chandrasekhar models, the violent merger model (Merger) provides the
best match to the observations, including the early rise and color
evolution. In detail, however, this model does not match the early wiggles
in the light curve, has consistently redder colors than \abc, and requires
$t_\mathrm{exp} \approx 2.3$\,days prior to $t_0$. As such, we postulate
that \abc\ is not the result of a violent WD--WD merger.

For clarity, of the two pure deflagration models only N5def is shown;
however, the evolution of N1600Cdef is very similar. While the pure
deflagration models produce the bluest colors at early times, they are
underluminous at times $>t_0 + 4\,\mathrm{days}$ and already rapidly
evolving toward the red at $\sim$$t_0 + 7$ (\abc\ exhibits a nearly constant
$(B-V)_0$ color for $\sim$19\,days after $t_0$). Thus, we conclude that
\abc\ is not compatible with pure deflagrations.

Of the $M_\mathrm{Ch}$ models, the W7 model better matches the
observations, as the N100 model features a faster rise and higher luminosity
than what is observed. We explore delayed-detonation models in further detail
below.

\subsection{Interaction with Nearby, Unbound Material}

To model SN\,2011fe, \citet{2014MNRAS.441..532D} examined pulsational
delayed-detonation (PDD) models as an explanation for some SNe Ia. Briefly,
PDD models differ from ``standard'' delayed-detonation (DD) models in that
the expansion of the WD during the initial deflagration phase leads to the
release of unbound material. Following this pulsation, the bound material
contracts, eventually triggering a subsequent
detonation.\footnote{\citet{2014MNRAS.441..532D} note that the deflagration
and detonation in their PDD models are artificially triggered.} An important
consequence of this progression for PDD models is that the unbound material
expands and avoids burning, unlike DD models that typically leave no unburnt
material. This results in significantly more carbon in the outer layers of
the SN ejecta \citep{2014MNRAS.441..532D}.

\citet{2014MNRAS.441..532D} find that DD models are universally faint and
red at early times, $\sim$24--48\,hr after explosion, while the PDD models
exhibit a faster rise and bluer colors. Briefly, this occurs in the PDD
scenario because the collision with the unbound material surrounding the WD
heats the outer layers of the SN ejecta. Importantly, the PDD models are
nearly indistinguishable from DD models around peak and post-peak.

\begin{figure}[]
  \centering
  \includegraphics[width=3.35in]{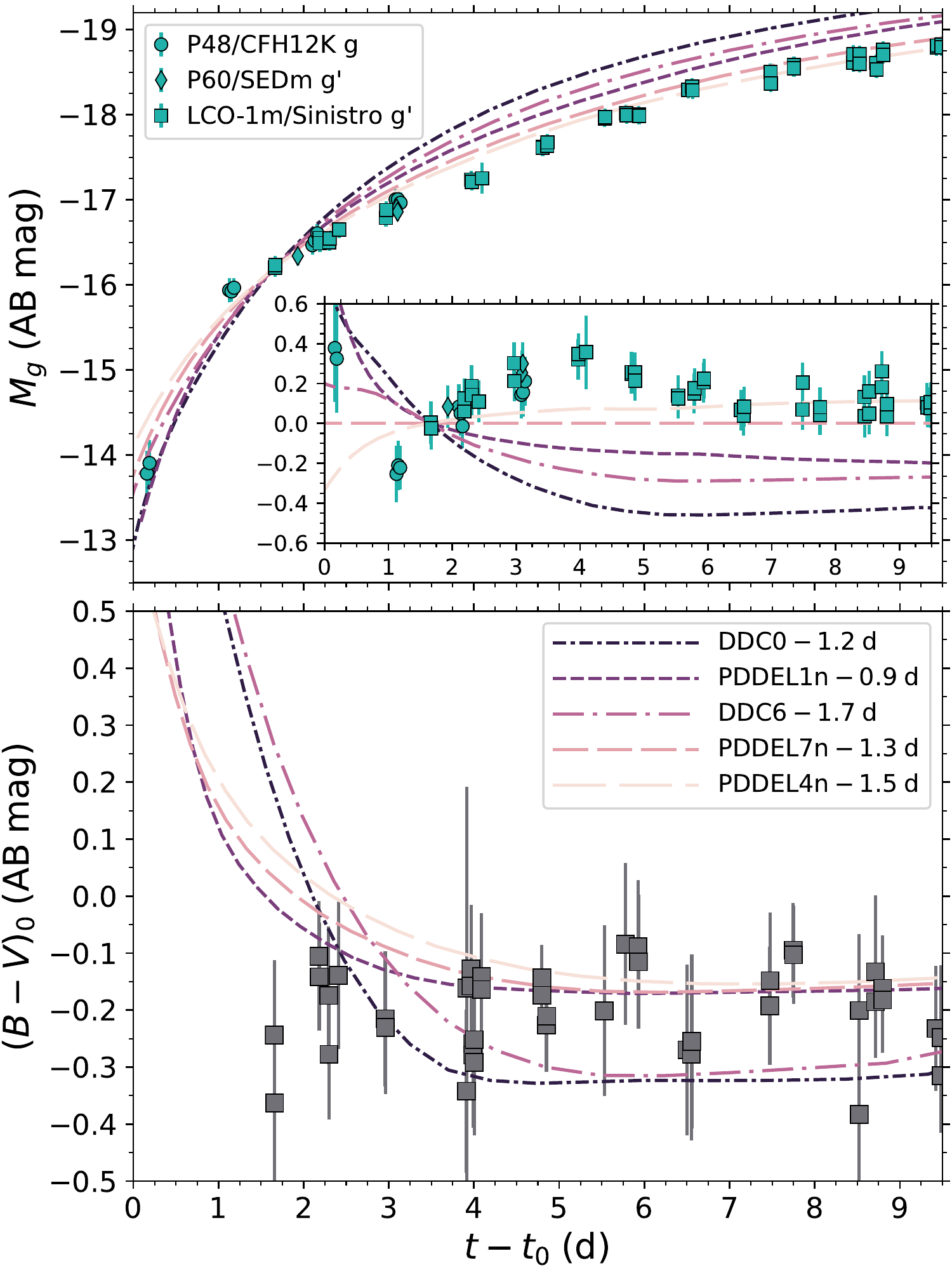}
  \caption{
  Same as Figure~\ref{fig:noebauer}, but featuring the $g$-band light curve
  and the DD and PDD models from \citet{2014MNRAS.441..532D}. The bottom
  panel legend lists the models in order of decreasing $M_\mathrm{Ni}$ from
  top to bottom. DD models (labeled as DDC to match the nomenclature of
  \citealt{2014MNRAS.441..532D}) are shown as dot-dashed lines, while PDD
  models (labeled as PDDEL) are shown as dashed lines. The inset in the top
  panel shows the residuals relative to the PDDEL7n model. The PDD models
  provide a better match to the observations.}
  \label{fig:dessart} 
\end{figure}

In Figure~\ref{fig:dessart} we compare photometric observations of \abc\ to
the high-$^{56}$Ni yield DD and PDD models presented in
\citet{2014MNRAS.441..532D}.\footnote{Available at
\url{https://www-n.oca.eu/supernova/snia/snia_ddc_pddel.html}.} DD and PDD
models with $M_\mathrm{Ni}/M_\odot \la 0.7$ and 0.5, respectively, fail to
match the early luminosity and blue colors of \abc. The PDD models provide a
better match to the observations than the DD models. The PDD models evolve
more rapidly toward blue colors and provide a better match to the $g$-band
flux in the days after explosion. Furthermore, unlike the DD models, the PDD
models exhibit strong \ion{C}{2} lines that gradually disappear in the
$\sim$1\,week after explosion \citep{2014MNRAS.441..532D}. In detail, the
early wiggles in the \abc\ light curve are not matched by the PDD models,
which exhibit more smooth variations. Furthermore, while the focus of this
study is the early evolution of \abc, the PDD models evolve more rapidly to
the red post-peak than the observations. Nevertheless, the PDD models
presented in \citet{2014MNRAS.441..532D} provide several attractive
explanations for the unusual features in the early behavior of \abc. Small
adjustments to the PDD models (e.g., additional $^{56}$Ni mixing, which
\citeauthor{2014MNRAS.441..532D} only explore for DD models) may better
match \abc.

\subsection{Strong $^{56}$Ni Mixing in the SN Ejecta}
\label{sec:Ni_mixing}

Having examined other possibilities, we now consider whether the early
evolution of \abc\ can be explained simply by invoking strong mixing in the
SN ejecta. Strong mixing leads to a faster initial rise, as well as a more
rapid evolution toward blue colors.

\begin{figure}[]
  \centering
  \includegraphics[width=3.35in]{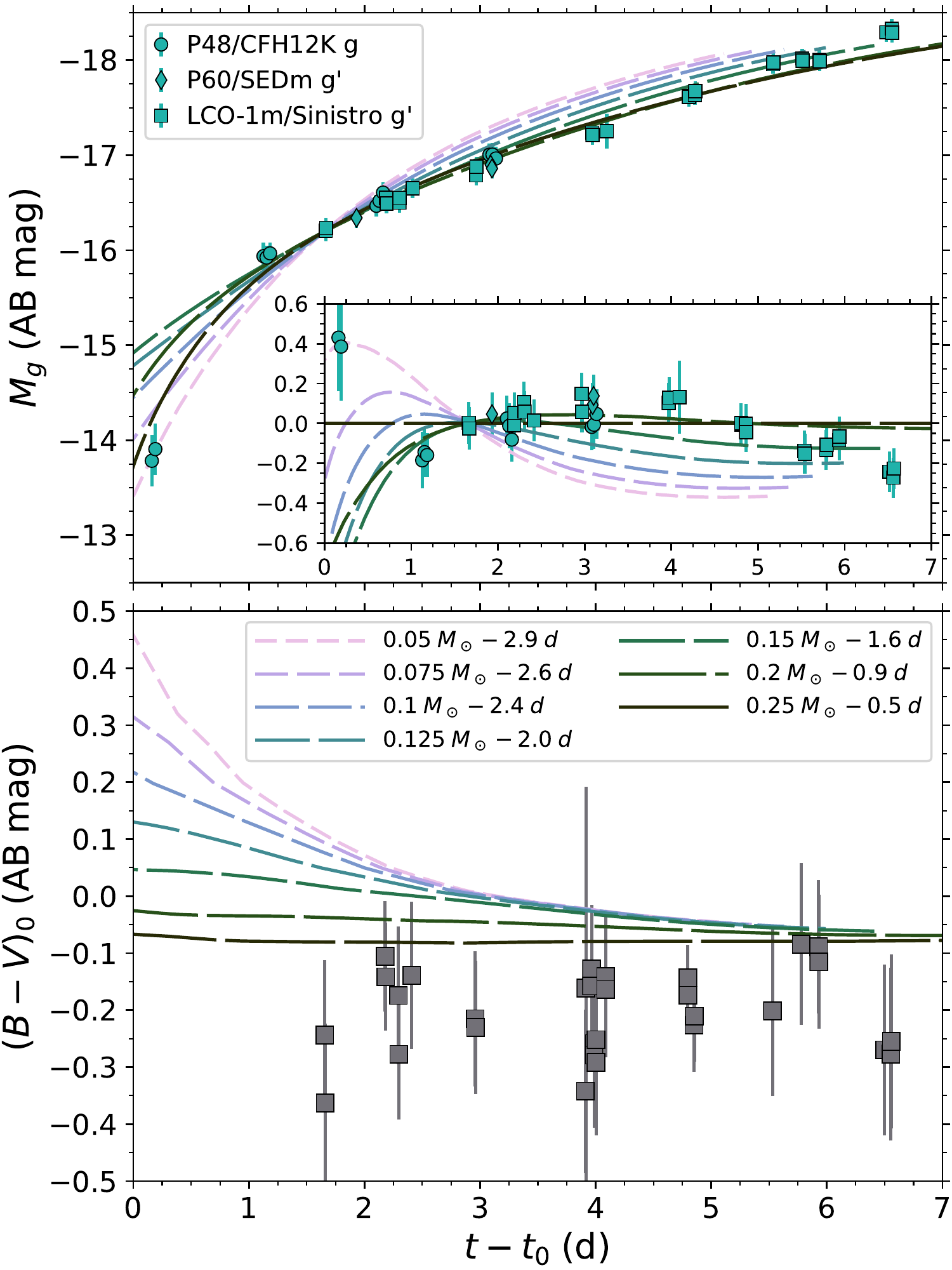}
  \caption{
  Same as Figure~\ref{fig:noebauer}, but featuring the $g$-band light curve
  and models from \citet{2016ApJ...826...96P}. The amount of $^{56}$Ni
  mixing in the SN ejecta increases from the light, short-dashed lines to
  the dark, long-dashed lines. Unlike Figure~\ref{fig:dessart}, each model
  features the same $M_\mathrm{Ni}$, while the model names reflect the
  boxcar widths used to approximate the effects of mixing in the ejecta (see
  \citealt{2016ApJ...826...96P}). The top panel inset shows the residuals
  relative to the $0.25\,M_\odot$ model. The observations are best matched
  by the $0.25\,M_\odot$ model, i.e. the model with the most significant
  mixing.}
  \label{fig:piro}
\end{figure}

Figure~\ref{fig:piro} compares the models from \citet{2016ApJ...826...96P}
to \abc. The \citeauthor{2016ApJ...826...96P} models employ a piston-driven
explosion to explode a single WD progenitor model. As the piston explosion
does not result in any nucleosynthesis, the distribution of $^{56}$Ni in the
ejecta must be prescribed by hand, which enables a study of the effects of
mixing on the resulting SN emission. Each model employs a fixed $0.5\,\sm$
of $^{56}$Ni that has been distributed throughout the ejecta via boxcar
averaging (see their Figure~1). The resulting light curves are synthesized
using the SuperNova Explosion Code (\texttt{SNEC};
\citealt{2015ApJ...814...63M}), as shown in Figure~\ref{fig:piro}. Broadly
speaking, the results can be summarized as follows: SNe with strong mixing
exhibit a rapid rise and quickly develop blue colors, whereas models where
the $^{56}$Ni is confined to the innermost layers of the ejecta remain faint
for days after explosion and feature a gradual color evolution from the red
to the blue. The model with the strongest mixing (dark long-dashed line in
Figure~\ref{fig:piro}) best matches the observations of \abc. This is the
only model we have found that exhibits a flat $(B-V)_0$ color evolution in
the days after explosion; however, in detail this model is too red relative
to the observations of \abc.

While the models from \citet{2016ApJ...826...96P} provide a good match to
the optical photometric evolution of \abc, they consistently overpredict the
flux in the UV. They also overpredict the photospheric velocity of \abc\ by
$\sim$2--3000$\; \mathrm{km \; s}^{-1}$. Furthermore, the simple gray
opacities in SNEC likely produce a faster rise and bluer colors than the
more detailed treatments employed in \citet{2014MNRAS.441..532D} and
\citet{2017MNRAS.472.2787N}.

Both the PDD models and ejecta-mixing models show discrepancies with some
early observations of \abc. Nevertheless, we conclude that one, or both, of
these scenarios, which feature qualitatively similar predictions, is the
most likely explanation for \abc. Indeed, it may be the case that the
typical sequence of photometric and spectroscopic observations of young SNe
Ia can never distinguish between these two possibilities
\citep{2017MNRAS.472.2787N}.

\section{The Emerging Sample of Young \sneia}

The proliferation of high-cadence, time-domain surveys has led to several
SNe Ia being discovered within $\sim$2\,days of first light in roughly the
past decade. Observations probing the early evolution of these SNe allow us
to place unique constraints on their progenitor systems and the
corresponding explosion physics. This has revealed diversity in the earliest
epochs after explosion, and that commonly used SN Ia templates do not match
observations at these phases (e.g., \citealt{2012ApJ...744...38F}).

In \S\ref{sec:first_light} we compared our early observations of \abc\ to
SN\,2011fe, which has the most comprehensive observations of the young SN Ia
sample. Given its normal spectroscopic and photometric evolution, SN\,2011fe
has been adopted as a standard for the early evolution of SNe Ia in many
studies. While a detailed quantitative analysis is beyond the scope of this
study, a qualitative examination of very young SNe Ia that otherwise exhibit
normal spectra and evolution at peak and post-peak\footnote{This definition
excludes iPTF\,14atg, which was shown to be subluminous with SN\,2002es-like
spectra \citep{2015Natur.521..328C}.} reveals considerable diversity. In
other words, at early times SN\,2011fe may not be the norm.

For SN\,2011fe the initial rise is well described by a $t^2$ power law, the
$(B - V)_0$ colors evolve from the red to the blue in the $\sim$1\,week
after explosion, and the \ion{C}{2} present in the initial spectra is weak
\citep{2011Natur.480..344N,2012ApJ...752L..26P,2016ApJ...820...67Z}. In
contrast, \abc\ exhibits a near-linear rise in flux, the $(B - V)_0$ colors
are blue and roughly constant, and the \ion{C}{2} absorption is strong.
Examining just these three qualitative features, SN\,2009ig is well matched
to SN\,2011fe \citep{2012ApJ...744...38F}, while SN\,2013dy
\citep{2013ApJ...778L..15Z}, SN\,2017cbv \citep{2017ApJ...845L..11H}, and
\abc\ all bear a striking resemblance. SN\,2012cg, on the other hand, is
intermediate to these two groups, with weak \ion{C}{2} and a relatively
shallow early rise, like SN\,2011fe, but blue $(B - V)_0$ colors, like \abc\
\citep{2012ApJ...756L...7S,2016ApJ...820...92M}. SN\,2014J is intermediate
in the other direction in that it exhibits a near-linear rise
\citep{2014ApJ...783L..24Z,2015ApJ...799..106G}, but the color evolution is
very similar to that of SN\,2011fe
\citep{2014ApJ...788L..21A}.\footnote{\ion{C}{2} is not detected in the
spectra of SN\,2014J \citep{2014ApJ...784L..12G,2014ApJ...783L..24Z}, though
the earliest spectra of SN\,2014J were obtained at a much later phase than
the other SNe discussed here. \citet{2015ApJ...798...39M} find evidence for
\ion{C}{1} in the NIR spectra of SN\,2014J, but this detection cannot
constrain the relative strength of \ion{C}{2} and \ion{Si}{2} shortly after
explosion.} That these early observations cannot be easily separated into
two distinct groups suggests that it is unlikely that a single physical
mechanism drives the diversity of SNe Ia at early times.

In the case of SN\,2012cg and SN\,2017cbv it has been argued that the early
blue optical colors are indicative of interaction between the SN ejecta and
a binary companion \citep{2016ApJ...820...92M,2017ApJ...845L..11H}.
SN\,2017cbv is particularly remarkable in that the observations presented in
\citet{2017ApJ...845L..11H} show a clearly resolved bump in the $U$, $B$,
and $g'$ bands in the $\sim$5\,days after explosion. In
\citeauthor{2017ApJ...845L..11H} it is found that the bump can be explained
via the combination of ejecta--companion interaction and the normal
evolution of a SN Ia. A challenge for this model, similar to \abc\ (see
\S\ref{sec:companion}), is that it significantly overpredicts the UV
brightness of the SN compared to what is observed. Indeed, in the case of
SN\,2012cg, SN\,2017cbv, and \abc\ the $\mathrm{UV} - \mathrm{optical}$
colors are significantly redder than those observed in iPTF\,14atg. It is
argued in \citet{2017ApJ...845L..11H} that several model assumptions,
including (i) ideal blackbody emission, (ii) a constant opacity, (iii) a
simple power-law density profile for the ejecta, and (iv) spherical
symmetry, may be incorrect, which could reconcile the discrepancy with the
UV observations. Above, we argued for interaction with diffuse, unbound
material and strong $^{56}$Ni mixing as a possible explanation for \abc, and
indeed \citet{2017ApJ...845L..11H} consider these possibilities for
SN\,2017cbv as well. Separately, several arguments against companion
interaction for SN\,2012cg are presented in \citet{2016arXiv161007601S}.

Ultimately, there are arguments in favor of and against each of the
possibilities to model the early emission from SNe Ia. Moving forward, more
detailed models and simulations are needed to properly explain the observed
diversity. No matter the correct explanation for the early behavior of
SN\,2013dy, SN\,2017cbv, and \abc, the strong similarities between these
events suggest that they may reflect a common physical origin.

\section{Conclusion}
\label{sec:conclusion}

We have presented observations of the extraordinarily early discovery of the 
normal SN Ia \abc. Our fast-response follow-up 
campaign allowed us to draw the following conclusions:

\begin{enumerate} 
    
    \item Extrapolation of the early light curve shows that the initial
    detection of \abc\ occurred only $0.15\pm_{0.07}^{0.15}$~days after the
    time of first light, $t_0$.

    \item We find no evidence for detectable signatures of SN shock cooling
    or the collision of the SN ejecta with a non-degnerate binary companion.

    \item Assuming that $v_\mathrm{ph} \propto t^{-0.22}$, then
    $t_\mathrm{exp} \approx t_0$. A short dark phase, as this implies, is
    likely the result of either strong $^{56}$Ni mixing or interaction of
    the SN ejecta with nearby, unbound material.

    \item The strong and short-lived carbon features seen in the earliest
    spectra of \abc\ can only be explained if there is incomplete burning.
    The pulsational delayed-detonation models presented in
    \citet{2014MNRAS.441..532D} produce \ion{C}{2} absorption that is as
    strong as \ion{Si}{2} at very early phases.

    \item In contrast to SN\,2011fe, $(B - V)_0$ is $\sim$0.5\,mag bluer for
    \abc\ at $t \approx -16$\,days. Furthermore, the $(B - V)_0$ colors of
    \abc\ show no evolution over the first $\sim$7\,days of observations.

    \item Finally, we show that the early light-curve evolution and colors
    of \abc\ are best matched by the pulsational delayed-detonation models
    of \citet{2014MNRAS.441..532D} and the ejecta-mixing models of
    \citet{2016ApJ...826...96P}.
 
\end{enumerate}
Taken together, these observations suggest that the early emission from
\abc\ is due to either the collision of the SN with nearby, unbound material
and/or that there is significant mixing of $^{56}$Ni in the SN ejecta. The
PDD models from \citet{2014MNRAS.441..532D} are particularly attractive for
explaining \abc, because they produce strong \ion{C}{2} absorption at early
times. In the future, it would be useful to investigate more detailed PDD
models that incorporate strong $^{56}$Ni mixing to see whether they better
replicate the observations of \abc, as it is otherwise difficult to
distinguish between these two scenarios.

Extremely early observations of young SNe provide a ``smoking gun'' to probe
the mixing level in the ejecta, which, in turn, is a result of the explosion
mechanism. Wide-field, high-cadence surveys, such as the Zwicky Transient
Facility \citep{2016PASP..128h4501B} and ATLAS
\citep{2011PASP..123...58T,2013RSPTA.37120269T}, will discover a large
number of very young SNe over the next few years, allowing us to extend our
studies beyond single objects. While the sample of extremely young SNe Ia
will grow by more than an order of magnitude, the detection of shock
breakout cooling and ejecta--companion interaction will prove challenging.
Given the diminutive size of WDs, the thermal emission following shock
breakout can only be detected to $\sim 10\,\mathrm{Mpc}$ on 1\,m class
telescopes. Furthermore, only $\sim$10\% of single-degenerate progenitors
are expected to give rise to detectable emission following the collision of
the SN ejecta with the binary companion \citep{2010ApJ...708.1025K}. Despite
these limitations, this study of \abc\ shows that the early detection of SNe
Ia may probe explosion physics by measuring the amount of mixing in the SN
ejecta. Moving forward, a large sample of such objects will enable strict
constraints on the proposed explosion mechanisms for SNe Ia.

Finally, we close by emphasizing the importance of fast-response photometric
and spectroscopic follow-up campaigns. Without the early recognition of the
youth of this SN and the associated follow-up, much of the analysis
presented herein would not have been possible. The ability to trigger such
observations is essential to improve our physical understanding of SNe Ia.

\acknowledgements

This study has benefited from the suggestions of an anonymous
referee. We are deeply grateful to R.~C.~Thomas for indulging a slew of
questions regarding \ion{C}{2} in SNe. Similarly, it is our pleasure to buy a
beer for R.~Amanullah and U.~Feindt for discussions regarding SALT2.

This paper utilizes the LCO infrastructure for rapid and regular monitoring
of SNe, and we thank S.~Valenti and I.~Arcavi for their development efforts.
Figure~\ref{fig:branch_vel} would not have been possible without S.~Blondin
generously sharing the data from \citet{2012AJ....143..126B}.

Much of the analysis presented herein would not have been possible without
the help of several observers. We thank M.~West for taking the first
spectrum of \abc\ as a ToO on the DCT. We also thank P.~GuhaThakurta,
E.~C.~Cunningham, K.~A.~Plant, H.~Jang, and J.~Torres for executing a Keck
ToO as part of the UC/Caltech partnership, and also the Gemini service
observers for executing our ToO observations. Additionally, J.~Cohen,
N.~Suzuki, V.~Ravi, R.~Walters, A.~Ho, H.~Vedanthamand, K.~De, and L.~Yan
helped obtain data for this paper.

AAM is funded by the Large Synoptic Survey Telescope Corporation in support
of the Data Science Fellowship Program. YC acknowledges support from a
postdoctoral fellowship at the eScience Institute, University of Washington.

DAH, CM, and GH are supported by NSF-1313484.

The Intermediate Palomar Transient Factory project is a scientific collaboration among the California Institute of Technology, Los Alamos National Laboratory, the University of Wisconsin, Milwaukee, the Oskar Klein Center, the Weizmann Institute of Science, the TANGO Program of the University System of Taiwan, and the Kavli Institute for the Physics and Mathematics of the Universe. This work was supported by the GROWTH project funded by the National Science Foundation under Grant No 1545949. Part of this research was carried out at the Jet Propulsion Laboratory, California Institute of Technology, under a contract with the NASA. This work makes use of observations from the LCO network. These results made use of the Discovery Channel Telescope at Lowell Observatory. Lowell is a private, nonprofit institution dedicated to astrophysical research and public appreciation of astronomy and operates the DCT in partnership with Boston University, the University of Maryland, the University of Toledo, Northern Arizona University, and Yale University. The upgrade of the DeVeny optical spectrograph has been funded by a generous grant from John and Ginger Giovale. Based on observations made with the Nordic Optical Telescope, operated by the Nordic Optical Telescope Scientific Association at the Observatorio del Roque de los Muchachos, La Palma, Spain, of the Instituto de Astrof\'isica de Canarias.

\facility{DCT, Gemini:Gillett, Hale, Keck:I, Keck:II, LCOGT, PO:1.2\,m, PO:1.5\,m, NOT, VLT, \textit{Swift}, OANSPM:HJT}

\software{\texttt{PTFIDE} \citep{2017PASP..129a4002M}, \texttt{FPipe}, \texttt{SALT2}, \texttt{SNID}, \texttt{lcogtsnpipe}, \texttt{UVOTSOURCE}}.

\appendix

\section{Photometric Light Curves}

\begin{figure}[ht]
  \centering
  \includegraphics[width=7in]{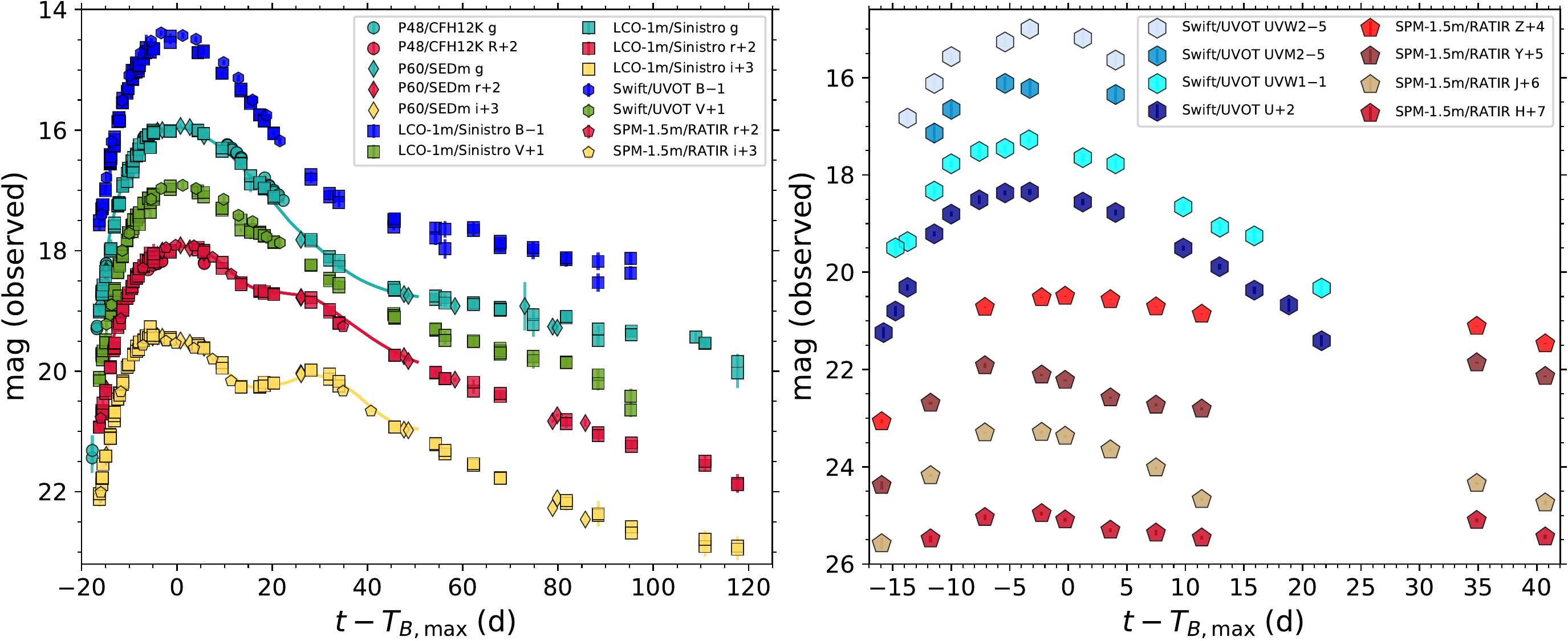}
  \caption{
  UV, optical, and NIR light curves for \abc. \textit{Left}: $BVgri$ light
  curves from the P48, P60, LCO-1m, \textit{Swift}, and SPM-1.5m telescopes.
  The solid lines represent the best-fit model from SALT2 (see
  \S\ref{sec:classification}). \textit{Right}: UV and NIR light curves from
  the \textit{Swift} and SPM-1.5m telescopes, respectively. For both panels
  each light curve is represented with a different color, while the
  different symbols correspond to different instruments, as detailed in the
  legends. The legends also list offsets applied to each light curve.
  Photometry is shown in the AB mag system, with the exception of the $BV$
  bands, which are shown in the Vega system. These light curves have not
  been corrected for line-of-sight extinction.}
  \label{fig:ugly_LC}
\end{figure}

The full photometric light curves of \abc\ are shown in Figure~\ref{fig:ugly_LC}.

\end{document}